\documentclass[aps,amsmath,amssymb,nofootinbib,superscriptaddress,showpacs,
floatfix,twocolumn,prb,10pt]{revtex4-1}

\usepackage{latexsym}
\usepackage{graphicx}
\usepackage{times,psfrag,subfigure}
\usepackage{amsmath}
\usepackage{dcolumn}
\usepackage{bm,bbm}       
\usepackage{color}
\usepackage{latexsym,amsmath,amssymb,bm,euscript}
\usepackage{chngcntr}
\counterwithin*{paragraph}{subsection}

\def\Tr{\mathop{\mathrm{Tr}}}

\newcommand{\beq}{\begin{equation}}
\newcommand{\eeq}{\end{equation}}
\newcommand{\beqarray}{\begin{eqnarray}}
\newcommand{\eeqarray}{\end{eqnarray}}
\newcommand{\Ref}[1]{Ref.~\onlinecite{#1}}  
\newcommand{\Sec}[1]{Sec.~\ref{#1}}  
\newcommand{\eq}[1]{Eq.~(\ref{#1})}  
\newcommand{\fig}[1]{Fig.~\ref{#1}}  
\newcommand{\Fig}[1]{Fig.~\ref{#1}}

\newcommand{\Z}{\mathbb{Z}} 

\definecolor{DarkRed}{RGB}{193,40,40}

\graphicspath{{Graphics/}}

\begin{document}

\allowdisplaybreaks

\title{A search for correlation-induced adiabatic paths between distinct topological insulators}

\date{\today}

\author{Johannes S. Hofmann}
\email{jhofmann@physik.uni-wuerzburg.de} 
\affiliation{Institut f\"ur Theoretische Physik und Astrophysik,
Universit\"at W\"urzburg, Am Hubland, D-97074 W\"urzburg, Germany}
\affiliation{Department of Condensed Matter Physics, Weizmann Institute of Science,
	Rehovot 76100, Israel}

\author{Fakher F. Assaad}
\affiliation{Institut f\"ur Theoretische Physik und Astrophysik,
             Universit\"at W\"urzburg, Am Hubland, D-97074 W\"urzburg, Germany}
\affiliation{W\"urzburg-Dresden Cluster of Excellence ct.qmat,
             Universit\"at W\"urzburg, Am Hubland, D-97074 W\"urzburg, Germany}

\author{Raquel Queiroz}
\affiliation{Department of Condensed Matter Physics, Weizmann Institute of Science,
	Rehovot 76100, Israel}

\author{Eslam Khalaf}
\affiliation{Department of Physics, Harvard University, Cambridge, MA 02138}

%TODO Fix placing on indices!
%ARXIV Monday 2+eplsion EST

\begin{abstract}
Correlations in topological states of matter provide a rich phenomenology, including a reduction in the topological classification of the interacting system compared to its non-interacting counterpart. This happens when two phases that are topologically distinct on the non-interacting level become adiabatically connected once interactions are included. Here, we use a numerically exact quantum Monte Carlo method to study such a topological reduction. We consider a 2D charge-conserving analog of the Levin-Gu superconductor whose classification is reduced from $\Z$ to $\Z_4$. We may expect any symmetry-preserving interaction that leads to a symmetric gapped ground state at strong coupling, and consequently a gapped symmetric surface, to be sufficient for such reduction. Here, we provide a counter example by considering an interaction which (i) leads to a symmetric gapped ground state at sufficient strength and (ii) does not allow for any adiabatic path connecting the trivial phase to the topological phase with $w = 4$.
The latter statement is established by numerically mapping the phase diagram as a function of the interaction strength and a parameter tuning the topological invariant. 
Instead of the adiabatic connection, the system exhibits an extended region of spontaneous symmetry breaking separating the topological sectors. Upon the inclusion of frustration, the size of this long-range ordered region is reduced until it gives way to a first order phase transition. 
Within the investigated range of parameters, there is no adiabatic path deforming the formerly distinct free fermion states into each other. We conclude that an interaction which trivializes the surface of a gapped topological phase  is a necessary but not  sufficient condition  to establish an adiabatic path 
within the interaction-reduced classification. In other words, the class of interactions which trivializes the surface is \emph{different} (likely larger) from the class of interactions which establishes an adiabatic connection in the bulk.

\end{abstract}

\date{\today}

\maketitle

\section{Introduction}

Recent years have witnessed an intense research effort to understand topological phases of matter \cite{hasan10_rev,qi11,chiu_review15,ludwig15}.
Symmetry-protected topological (SPT) phases are described by equivalence classes of phases under symmetric adiabatic deformation. This means that two SPTs belonging to the same phase can be deformed to each other without closing the bulk gap or breaking the protecting symmetries, whereas two distinct SPT phases cannot. SPTs protected by internal symmetries such as time-reversal or particle-hole symmetry have been extensively studied for free-fermion systems \cite{schnyder08,ryu10}. These are all characterized by the existence of gapless anomalous surface states whose existence is a direct consequence of the bulk topology, a phenomenon known as bulk-boundary correspondence.

The inclusion of interactions can modify the topological character of free-fermion SPTs in at least three different ways: (i) The spontaneous breaking of protecting symmetry can lead to the disappearance of surface states and consequently alter the topological classification. (ii) Correlations can induce topological order that is characterized by long-range entanglement as in fractional quantum Hall states \cite{Tsui82,stromer99}, fractional topological insulators \cite{levin09}, or quantum spin liquids \cite{balents-10,Lee08,Yamashita10,sachdev08}. These states do not have a non-interacting analog. (iii) The free-fermion topological classification may be reduced due to the existence of symmetric adiabatic paths in the space of interacting Hamiltonians connecting states that are disconnected at the non-interacting level \cite{fidkowski10a,tang12,yao13,morimoto15,qi13,queiroz16,turner11,fidkowski11,gu14,wang14e,senthil15,vishwanath13}.

The first example of a single-particle topology reduction was considered by Fidkowski and Kitaev in Ref.~\onlinecite{fidkowski10a}. In this work, the authors study a spinless superconductor in one dimension (Kitaev chain), with spinless time-reversal symmetry $\mathcal{T}^2=+1$, representing class BDI. They have constructed an explicit interaction which preserves the symmetry and has a unique and symmetric ground state, that yet gaps out 8 topological Majorana boundary modes and allows an adiabatic connection between bulk states whose winding number differ by 8. This implies a reduction of the non-interaction classification from $\mathbb{Z}$ to $\mathbb{Z}_8$. Later on, this result has been generalized to different symmetry classes and higher dimensions \cite{tang12,yao13,morimoto15,qi13,queiroz16,turner11,fidkowski11,gu14,wang14e,senthil15,vishwanath13}. Whereas the early work by Fidkowski and Kitaev \cite{fidkowski10a} used the simple properties of the 1D model to show explicitly that two phases differing by winding of 8 can be adiabatically connected, most of the investigations in higher dimensional systems relied on bulk-boundary correspondence to argue that the existence of an interaction which symmetrically gaps out the topological boundary modes is sufficient to establish the collapse of the non-interacting classification. These approaches employed several arguments such as studying the possibility of gapping surface states  \cite{tang12,yao13,morimoto15} or 0D defects that follow from dimensional reduction \cite{qi13,queiroz16} as well as investigating the signatures of these boundary states in the entanglement spectrum \cite{turner11,fidkowski11}. Other bulk-based approches include studying the braiding statistics arising from gauging the symmetry \cite{gu14} or group cohomology \cite{wang14e}. 

In spite of this body of work, which provided a comprehensive answer to the question of the general interaction-reduced classification for topological phases protected by internal symmetries, only a few works \cite{He16} addressed the fate of a given topological phase in the presence of a specific symmetric interaction. In particular, we would like to investigate the question whether an interaction which symmetrically gaps out surface states at sufficient strength also enables an adiabatic symmetric path connecting two distinct non-interacting SPTs. In this work, we show that this is not generally  true. As we will illustrate in detail, the caveat is that the interaction results in transitions into symmetry-broken phases along the paths, preventing adiabaticity despite the fact that at strong enough interaction strength the ground state is symmetric.

To show this, we consider a simple model of four identical layers of topological insulators protected by charge conservation and an internal $\Z_2$ symmetry (a non-superconducting analog of the one considered by Levin and Gu \cite{gu14}) whose non-interacting $\Z$ classification is reduced to $\Z_4$. The non-interacting theory has an emergent $SU(4)$ symmetry corresponding to rotations among the four different flavors and is characterized by the topological winding number $w = 4$.  However, we consider an interaction of much lower symmetry which reduces this flavor rotation symmetry to $U(1)\times U(1)$ (while still preserving the symmetries protecting the topological classification). It is worth mentioning here that our approach differs from earlier works \cite{fidkowski11,ayyar15,He16} which considered highly symmetric interactions to avoid possible symmetry breaking. Here, we consider an interaction with very low symmetry precisely to show that symmetry breaking is unavoidable along the adiabatic path although the ground state is still symmetric at large enough interaction strength.

Starting from a microscopic model, we employ the projective auxiliary-field Quantum Monte Carlo method \cite{blankenbecler81,white89,assaad08_rev,alf_v1} to study its ground state phase diagram. We find that adiabatic connection between the $w = 4$ and $w = 0$ phases of the model is not possible due to the appearance of an extended region of spontaneous symmetry breaking that additionally separates
the single particle topological sectors. To overcome this problem, we add extra terms to the Hamiltonian to frustrate the long range order. At weak levels of frustration, the region of spontaneous symmetry breaking is reduced in size until it gives way to a first order phase transition at strong frustration that still blocks the adiabatic connection. As a result, we conclude that the interaction we considered, while sufficient for gapping out surface states, is insufficient for the existence of a symmetric adiabatic deformation between the trivial and non-trivial phases. This poses a counter example to the criteria derived by some of the authors in Ref.~\onlinecite{queiroz16} which is shown to be necessary but not sufficient.

This article is organized as follows. In \Sec{spt:model}, we define the two-dimensional microscopic model and discuss its symmetries.In \Sec{spt:MF}, we analyze possible mean-field scenarios with spontaneous symmetry breaking and identify the most dangerous channel. Additionally, we present the analytic solution in the limit of infinite interaction strength and find a unique, symmetric and gapped ground state. In \Sec{spt:method}, we briefly discuss the projective auxiliary-field Quantum Monte Carlo (QMC) method.
We present the numerical extracted phase diagram in \Sec{spt:results} that exhibits a region of spontaneous symmetry breaking that gives way to a first order phase transition at strong frustration. In \Sec{spt:discussion}, we conclude with a discussion of the implications of the phase diagram and suggest future avenues.

\section{Model \& symmetries}
\label{spt:model}

Here, we design a two-dimensional microscopic model that obeys an anti-unitary time-reversal symmetry (TRS) and a unitary $\mathbb{Z}_2$ symmetry \cite{queiroz16, qi13}. The latter can be implemented, e.g., as the conservation of the $z$ component of the spin $S^z$ modulo $2$. The topology of these free fermion model is given by a $\mathbb{Z}$ valued winding number $w$ that is related to the number of helical Dirac cones at the edge of the sample. In the presence of correlations, the classification is expected to be reduced from $\mathbb{Z}$ to $\mathbb{Z}_4$. We define a specific interaction term that should allow for adiabatic deformations of free fermion states whose winding number differ by (multiples of) $\Delta w=\pm4$.

\begin{figure}
	\begin{center}
		\includegraphics[width=0.995\columnwidth]{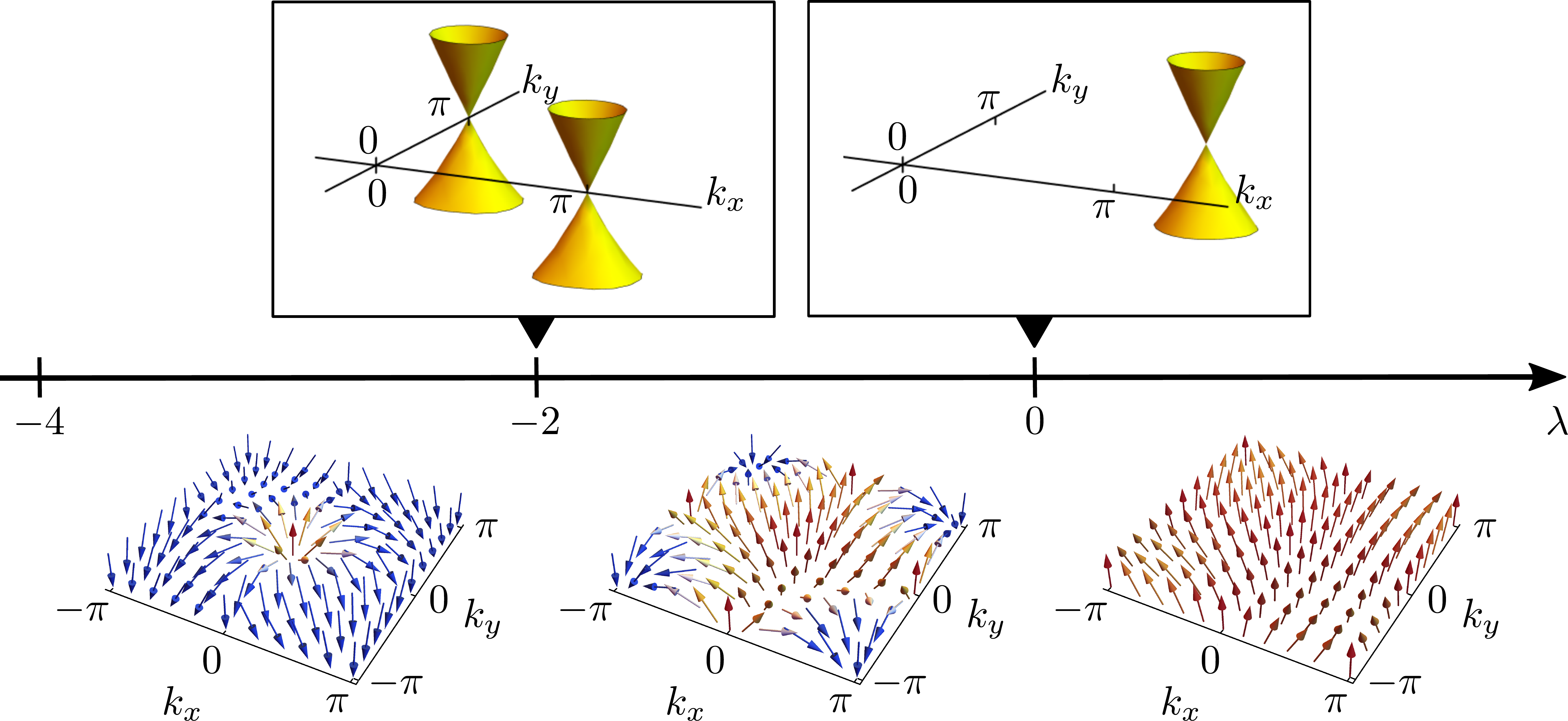}
	\end{center}
	\caption{\label{fig:topol}Graphical visualization of $\mathcal{H}_0$ showing the localization of the Dirac cone(s) for $\lambda=-2$ and $\lambda=0$. The 3D-vector plots depict the orientation of $\hat{\mathbf{d}}$ representing topological (left and middle) and trivial (right) insulators. For readability, we also color-coded the $z$-component.
	}
\end{figure}

We begin by introducing the free fermion part of the model $\mathcal{H}_0=\sum_{\mathbf{k} }\Psi_{\mathbf{k}}^{\dagger} H(\mathbf{k}) \Psi_{\mathbf{k}}$ which represents two copies of a Quantum-Hall system with opposite Chern number representing a topological insulator \cite{kane2005,bernevig2006,qi08a,liu08} which preserves the spin projection $\sigma_z$. We have
\begin{subequations}
	\label{Eq:H0Full}
	\begin{eqnarray}
	H(\mathbf{k})&=&\sin(k_x)\sigma_0\tau_x+\sin(k_y)\sigma_0\tau_y + m(\mathbf{k})\sigma_z\tau_z 
	\nonumber \\&&
	+ \Delta[\sin^2(k_x)+\sin^2(k_y)]\sigma_0\tau_0 \label{Eq:H_R_sector} \\
	m(\mathbf{k})&=&2+\lambda+\cos(k_x) + \cos(k_y) \label{Eq:mass} \, ,
	\end{eqnarray}
\end{subequations}
where $\Psi_{\mathbf{k}}^\dagger$ is the creation operator of a four-component spinor with momentum $\mathbf{k}$. The above Hamiltonian is block diagonal and we denote the sub-blocks by $H_\pm(\mathbf{k})$. They represent a Chern insulator with a $\pm 1$ Chern number \cite{haldane88a}. 
The energy scale is set by $t$ which will be used as the unit of energy ($t=1$) throughout the rest of this manuscript. \eq{Eq:H0Full}, corresponds to a gapped Dirac cone at each of the four time-reversal invariant momenta, with the gap dictated by $m(\mathbf{k})$. For the choice of parameters $\lambda=0$, $\lambda=-2$ or $\lambda=-4$ at least one of the Dirac cones remains gapless (compare with \Fig{fig:topol}) whereas any other value describes an insulating state. These points correspond to topological phase transitions. The last term in \eq{Eq:H_R_sector}, $\Delta = 0.25$ is used to break the particle-hole symmetry within $H_\pm(\mathbf{k})$.
We note that the dimensional reduction arguments used in Ref.~\onlinecite{queiroz16} relied on the existence of chiral symmetry in the continuum model. The symmetry is explicitly broken here following the aforementioned scheme of having as little symmetries as possible. Also in real materials, the particle-hole symmetry is an approximate low energy symmetry, unless one considers superconductors.

The Hamiltonian $H(\bf k)$ obeys a spurious $U(1)$ symmetry $\exp\{i\phi\sigma_z\}$ instead of the required  $\mathbb{Z}_2$  Ising symmetry  $R=\sigma_z$. It separates the $H_\pm(\bf k)$ sectors, where $\pm$ is given by the eigenvalues of $R$. Additionally, there is one independent, anti-unitary time-reversal symmetry $\mathcal{T}=\sigma_y \tau_y\,\mathcal{K}$. Here, $\mathcal{K}$ refers to the complex conjugation. Hence, the model is very closely related to the well know topological insulators (TIs). There, one may introduce spin-orbit coupling which breaks the spin conservation as long as the time-reversal symmetry is respected. 

In order to discuss the topology of the gapped states, it is useful to first focus on $H_+$. Defining the three-component vector $\mathbf{d}=(\sin(k_x),\sin(k_y), m(\mathbf{k}))$, the Chern number of $H_+$ takes the form \begin{align}
C_+=\frac{1}{4\pi}\iint_{\textrm{BZ}}\hat{\mathbf{d}} \cdot (\partial_{k_x} \hat{\mathbf{d}} \times \partial_{k_y} \hat{\mathbf{d}}) \end{align} 
with $\hat{\mathbf{d}} =\mathbf{d} /|\mathbf{d}|$ \cite{bernevig2006}. As it can be seen in \Fig{fig:topol}, the parameter $\lambda$ tunes the system from a trivial insulator ($\lambda>0$) through a semi-metal with a Dirac cone at $\mathbf{k}=(\pi,\pi)$ ($\lambda=0$) to a Chern insulator ($-2<\lambda<0$) with Chern number $+1$. At $\lambda=-2$, the system exhibits one Dirac cone at each $\mathbf{k}=(0,\pi)$ and $\mathbf{k}=(\pi,0)$.
The winding number of the full Hamiltonian is then given as $w=(C_+ - C_-)/2$ where $C_-=-C_+$ due to the time-reversal symmetry connecting the two sectors.
Note that small values of $\Delta$ only modify the energy of the bands and, as long as the band gap does not close, the wave functions do not change as it represents a (momentum dependent) chemical potential. Hence the topology is insensitive to (small) $\Delta$.

To study the topological reduction from $\mathbb{Z}$ to $\mathbb{Z}_4$, we introduce four copies of $\mathcal H_0$ labelled by an orbital index $o\in \lbrace A, B, C, D\rbrace$. The dimensional reduction scheme\cite{queiroz16} can be used to derive the form of the interactions that allow the adiabatic connection by introducing a lattice of zero-dimensional defects. Such defects inherit the bulk topology in the sense that they exhibit $n$ topologically degenerate zero modes where $n$ matches the topological invariant of the bulk\footnote{Note that due to the broken particle-hole symmetry, the topological states do not necessarily have zero energy. However, a renormalized chemical potential can only shift the energy but it cannot lift their degeneracy.}. For two-dimensional models, this is done by first realizing one-dimensional edge modes at a domain wall and secondly adding an oscillating mass term along this domain wall with appropriately chosen symmetries, such that each node of the mass term localizes zero modes. This construction in turn allows to derive an explicit interaction term which gaps those defects without breaking any symmetry.
Using this recipe, we design the following interaction
\begin{align}
	\label{Eq:HintFull}
	\mathcal{H}_{\textrm{int}}&=&U\sum_{\mathbf{i},\alpha=\pm} \Psi^{\dagger}_{\mathbf{i},A}\gamma_5 P_{\alpha} \Psi^{\phantom{\dagger}}_{\mathbf{i},B}  \Psi^{\dagger}_{\mathbf{i},D}\gamma_5 P_{\alpha} \Psi^{\phantom{\dagger}}_{\mathbf{i},C} + h.c.
\end{align}
with the projectors $P_{\alpha}=\frac{1}{2}\left( 1 + i \alpha \gamma_3 \gamma_4 \right)$ satisfying $P=P^2$, and the $\gamma$ matrices acting on the original Dirac components as $\gamma_{1,2}=\sigma_0\tau_{x,y}$, and  $\gamma_{3,4,5}=\sigma_{z,y,x}\tau_{z}$.
Note that the spurious $U(1)$ symmetry is $R=i\gamma_4\gamma_5$ and that the operators $R$, $\gamma_4$ and $\gamma_5$ form an $SU(2)$ algebra. Hence, the symmetry generates continuous rotations in the $\gamma_4\gamma_5$ plane. Using only $\gamma_5$ in the interaction terms breaks the $U(1)$ symmetry down to the required discrete $\mathbb{Z}_2$ symmetry that transforms $\gamma_5\rightarrow - \gamma_5$.
Physically, this interaction introduces correlated pair hopping of electrons between layers $A\rightarrow B$ and $D\rightarrow C$ while flipping the $R$ charge. This term does allow, e.g., two $R=+$ electrons being scattered into two $R=-$ ones such that the $R$ charge is only conserved modulo 2 which illustrates the $U(1)\rightarrow \mathbb{Z}_2$ reduction\footnote{The attentive reader might have realized that this correlated hopping actually changes $R$ by $4$. However, in the canonical ensemble, the total number of particles is conserved and at half-filling with an even number of fermions, the total $R$ charge itself has to be even and thus this reduces the $U(1)$ symmetry to $\mathbb{Z}_2$.}.  We note that this interaction term is not invariant under any rotation between the flavors since it singles out a very specific channel of coupling the flavors. Thus, out of the SU(4) flavor rotation of the non-interacting theory, only a U(1)$\times$U(1) symmetry remains corresponding to simultaneous phase rotations of $\Psi_{A}$ and $\Psi_{B}$ or $\Psi_{C}$ and $\Psi_D$.

\section{Mean-field theory \& atomic limit}
\label{spt:MF}

\begin{figure}
	\begin{center}
		\includegraphics[width=0.99\columnwidth]{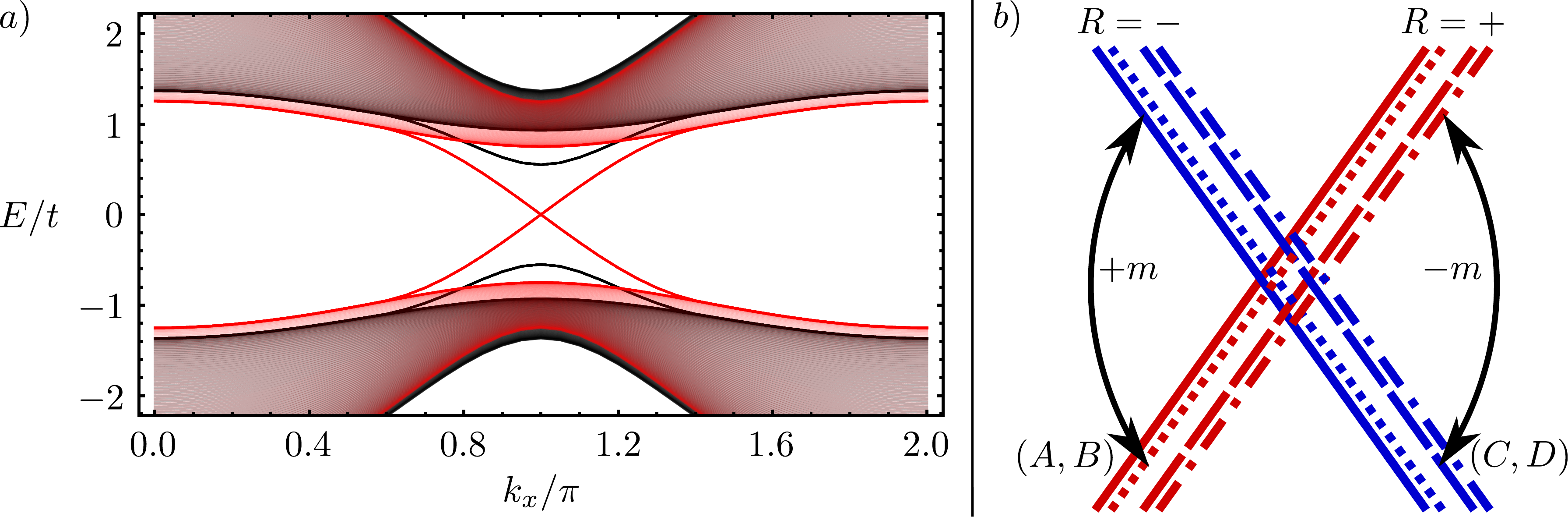}
	\end{center}
	\caption{\label{fig:MFoBC}(a) Spectrum of the model defined in \eq{Eq:H0Full} in ribbon geometry representing the symmetric (red) and the symmetry broken scenario (black) with an mean-field order parameter $m=0.6$ and  $\mathcal{H}_{\mathrm{MF}}=m\sum_{\mathbf{i}}M_{\mathbf{i}}^x$.
	(b) Sketch of some scattering processes induced by the order parameter to the (non-interacting) helical boundary modes. These terms will open a gap for $m\neq0$
	}
\end{figure}

Before presenting the method and the numerical results, let us develop an intuition on symmetric and symmetry breaking phases that the model presented in the previous section admits. Here, we discuss various mean-field scenarios and identify the channel which is most likely to exhibit symmetry breaking long-range order. We also argue that this phase may only be realized when the coupling strength is comparable or larger then the band gap $U_{c,1}<U$.
Additionally, we will solve the atomic limit analytically for $U\rightarrow\infty$. We find a unique and symmetric ground state that is gapped from the rest of the spectrum. Hence this state is also stable with respect to finite values of $U$. Accordingly, the symmetry broken phase is bounded both from below and above with $U_{c,1}<U<U_{c,2}$.
Interestingly, the states of the limiting cases all share the same symmetry such that an intermediate symmetry broken phase is not required. This provides another argument, complementary to the edge state analysis of \Ref{queiroz16}, on the existence of an adiabatic path and the according reduction of the topological classification.

\subsection{Weak interaction limit and Mean-field scenarios}

To discuss mean-field scenarios relevant at weak interactions, it is very useful to divide the four layers into two pairs, namely $(A,B)$ and $(C,D)$ and introduce the two pseudo-spin operators with $\beta=x,y$
\begin{subequations}
	\label{eq:pseudospin}
	\begin{eqnarray}
	S_{\mathbf{i}}^{1,\alpha,\beta}&=& 
	(\Psi^{\dagger}_{\mathbf{i},A} \Psi^{\dagger}_{\mathbf{i},B}) 
	\mu_\beta  \gamma_5 P_{\alpha}
	(\Psi_{\mathbf{i},A} \Psi_{\mathbf{i},B})^T
	\label{Eq:PseudoSpin1} \\
	S_{\mathbf{i}}^{2,\alpha,\beta}&=& 
	(\Psi^{\dagger}_{\mathbf{i},C} \Psi^{\dagger}_{\mathbf{i},D}) 
	\mu_\beta  \gamma_5 P_{\alpha}
	(\Psi_{\mathbf{i},C} \Psi_{\mathbf{i},D})^T
	\label{Eq:PseudoSpin2}
	\end{eqnarray}
\end{subequations}
where the Pauli matrices $\mu_\beta$ act on layer index within each pair. This allows us to rewrite the interaction as
\begin{equation}
\label{Eq:HintSumSq}
\mathcal{H}_{\textrm{int}}=\frac{U}{8} \sum_{\substack{\mathbf{i},\alpha=\pm,\\\beta=x,y}} \sum_{\zeta=\pm} \zeta\left( S_{\mathbf{i}}^{1,\alpha,\beta} + \zeta \, S_{\mathbf{i}}^{2,\alpha,\beta} \right)^2
\end{equation}
and shows the possibility to minimize the energy for $\zeta=-$, given that $U>0$, by the generation of pseudo-magnetic order in the $xy$-plane of
\begin{equation}
\label{Eq:MFchannel}
M^\beta_{\mathbf{i}}=\sum_{\alpha=\pm} \left( S_{\mathbf{i}}^{1,\alpha,\beta} - \, S_{\mathbf{i}}^{2,\alpha,\beta} \right) .
\end{equation}

Note that the operator $M^\beta_{\mathbf{i}}$ contain terms like $\Psi^{\dagger}_{\mathbf{i},A} \gamma_5 \Psi_{\mathbf{i},B}$. The required $R$ symmetry transforms $\gamma_5$ to $-\gamma_5$ and is therefore broken by the order parameter. Physically, $M^\beta_{\mathbf{i}}$ introduces single electron hopping processes, e.g., from layer $B$ to $A$, that flip the $R$ charge. We have illustrated this for the helical edge states in \Fig{fig:MFoBC}(b). This also points out that the order parameter generates a gapped edge spectrum.

Additionally, this order parameter anti-commutes with \eq{Eq:H0Full} such that it also introduces a gap for the bulk semi-metals. As \eq{Eq:HintSumSq} points out, the interaction is symmetric under rotation around the $z$-axis of the pseudo-spins, and hence the orientation of $M$ within the $xy$-plane is arbitrary and also breaks this symmetry spontaneously.

Without loss of generality, we choose the $x$ direction for the magnetization and introduce $\mathcal{H}_{\mathrm{MF}}=m\sum_{\mathbf{i}}M_{\mathbf{i}}^x$. In Fig.~\ref{fig:MFoBC}(a) we present the energy spectrum for $\mathcal{H}_0+\mathcal{H}_{\mathrm{MF}}$ with open boundary condition in $y$ direction. The solid black lines represent a non-zero mean-field expectation value $m>0$ and the red lines overlay the symmetric version ($m=0$). We can clearly observe the introduced gap of the former massless Dirac edge state.

To make the connection between this mean-field scenario and the phase diagram, let us discuss the limiting cases.
Keeping the interaction strength small, we expect stable Dirac cones for $\lambda\sim0$ and $\lambda\sim-2$ as the density of states at the Fermi level vanishes at half filling~\cite{sorella92,paiva05,meng10,sorella12,assaad13,toldin14}. The insulating states provide an intrinsic scale of energy, namely the band gap, such that the correlation should reach comparable strength before it leads to significant changes. Hence we expect that a symmetry broken phase, if at all, occurs at finite interaction strength $U>U_{c,1}$.

\subsection{Strong interaction limit}

The more interesting limiting case is the strongly interacting one with $U/t \gg 1$. Starting from the limit $t=0$, we can solve $\mathcal{H}_{\mathrm{int}}$ analytically, as the lattice sites completely decouple and we are left with a zero-dimensional problem. In the following, we calculate the full spectrum and show that there is a unique ground state. For readability, let us drop the position index $\mathbf{i}$ for the remainder of this analytic derivation. Note that $P_\pm$ act as projectors such that the Fock space can be decomposed into two separate blocks which have an identical spectrum. Hence it is sufficient to focus on one subspace, say $\mathcal{H}_+$. Let $\epsilon_i$ be an eigenvalue of $\mathcal{H}_+$ with degeneracy $g_i$. The full spectrum is then given by $\epsilon_i+\epsilon_j$ with degeneracy $g_i g_j$.

In the previous \Sec{spt:model}, we have chosen a basis for the $\gamma$-matrices in which $R$ is diagonal to remind the reader of popular QSH models. Here, however, it is more convenient to choose a different basis in which both $\gamma_5=\sigma_x\tau_z$ and $i\gamma_3\gamma_4=\sigma_x\tau_0$ are diagonal \footnote{Note that $\gamma_5$ and $P_\alpha$ commute with each other such that they may be diagonalized simultaneously. However, $\gamma_5$ anti-commutes with $R$ such that there is no basis which diagonalizes all three of them.}. Note that the local fermion degrees of freedom $\mathcal{H}_+$ within one layer $o$ is then fully classified by the eigenvalues $s_5=\pm$ of $\gamma_5$ such that $P_+\Psi_{o,s_5}=\Psi_{o,s_5}$ and $\gamma_5\Psi_{o,s_5}=s_5 \Psi_{o,s_5}$. This leads to the definition of four new spin operators:
\begin{eqnarray}
	\mathbf{S}_a & = & (\Psi_{A,+}^\dagger \Psi_{B,+}^\dagger) \boldsymbol{\mu}(\Psi_{A,+}^{\phantom{\dagger}} \Psi_{B,+}^{\phantom{\dagger}}) ^T \\
	\mathbf{S}_b & = & (\Psi_{A,-}^\dagger \Psi_{B,-}^\dagger) \boldsymbol{\mu}(\Psi_{A,-}^{\phantom{\dagger}} \Psi_{B,-}^{\phantom{\dagger}}) ^T \\
	\mathbf{S}_c & = & (\Psi_{C,+}^\dagger \Psi_{D,+}^\dagger) \boldsymbol{\mu}(\Psi_{C,+}^{\phantom{\dagger}} \Psi_{D,+}^{\phantom{\dagger}}) ^T \\
	\mathbf{S}_d & = & (\Psi_{C,-}^\dagger \Psi_{D,-}^\dagger) \boldsymbol{\mu}(\Psi_{C,-}^{\phantom{\dagger}} \Psi_{D,-}^{\phantom{\dagger}}) ^T \, ,
\end{eqnarray}
such that the Hamiltonian can be written as $\mathcal{H}_+=h_{ac}-h_{ad}-h_{bc}+h_{bd}$ where we used the shorthand notation $h_{ij}=U(S^+_i S^-_j + h.c.)$. Note that this has mapped \eq{Eq:HintFull} to a model with four sites on a ring ($a$-$c$-$b$-$d$) of spinful fermions. Each site can be empty, double occupied, or have a single fermion with up or down spin. The total Hilbert space is then $2^8=256$ dimensional. The Hamiltonian conserves the parity at each site\footnote{Actually, the local number of particles is conserved. However, in the following it is more useful to distinguish the parity only as both empty and double-occupied sites constitute a spin-singlet which effectively removes that site from the ring.} as well as the total $S^z$ spin component. Using the local parities, we can decompose the Hilbert space into 16 16-dimensional Hilbert spaces which can be studied as shown below.

We start by considering the subspaces where at least both $a$ and $b$ or both $c$ and $d$ sites are parity even. There are 7 16-dimensional subspaces with this property. Observe that every second site represents a spin singlet such that spin-flip processes on all bonds are prohibited. Hence, the Hamiltonian and all eigenvalues vanish. 

Next, we discuss the cases where only the $a$ or $b$ as well as only the $c$ or $d$ site is single occupied which occurs for 4 16-dimensional subspaces.  Then, only one bond operator, say $h_{ac}$, is non-zero. From the spin conservation follows that the sectors with maximal value of $|S^z|$ have a vanishing Hamiltonian as the spin-flip operators cannot act on those states. The $S^z=0$ subspace contains only two eigenstates with nonzero eigenvalues $\pm U$ given by $|\uparrow_a \downarrow_c\rangle \pm |\downarrow_a \uparrow_c\rangle$.

Third, we consider the subspace where exactly one parity is even (there are 4 such subspaces), e.g., site $b$, such that $\mathcal{H}_+=h_{ac}-h_{ad}$. Once more, the states with maximal $|S^z|=3/2$ are eigenstates of energy zero. In the $S^z=\pm 1/2$ subspace we find the Hamiltonian
\begin{equation}
	H_{1/2}= U
	\begin{pmatrix}
	0 & 1 & -1 \\
	1 & 0 & 0 \\
	-1 & 0 & 0
	\end{pmatrix}\, ,
\end{equation}
using the basis $\{ |\downarrow_a \uparrow_c \uparrow_d \rangle, |\uparrow_a \downarrow_c \uparrow_d \rangle, |\uparrow_a \uparrow_c \downarrow_d \rangle \} $. Here the eigenvalues are given by $\pm\sqrt{2}U$ and zero.

The last and most interesting subspace has only odd parity sites. Like before, the two states with $S^z=2$ are eigenstates with vanishing energy. In the $S^z=\pm 1$ sector we find eigenvalues of $0$ and $\pm2 U$. The wave function of the $-2U$ state is (from here on out, we drop the indicies for readability)
\begin{equation}
\phi_{1} = \frac{1}{2} \left( |\downarrow \uparrow \uparrow \uparrow \rangle - |\uparrow \downarrow \uparrow \uparrow \rangle - |\uparrow \uparrow \downarrow \uparrow \rangle + |\uparrow \uparrow \uparrow \downarrow \rangle \right) \,.
\end{equation}
Finally, for $S^z=0$ there are multiple states with vanishing energy, but only one state is associated with the eigenvalues $\pm2\sqrt{2}U$, respectively, where the ground state wave function is 
\begin{eqnarray}
\phi_{0} & = &  \frac{1}{2} \left( |\downarrow \downarrow \uparrow \uparrow \rangle + |\uparrow \uparrow \downarrow \downarrow \rangle \vphantom{\sqrt{2}} \right. \nonumber \\
&& + \left. \frac{1}{\sqrt{2}} [
   |\downarrow \uparrow \downarrow \uparrow \rangle
 - |\downarrow \uparrow \uparrow \downarrow \rangle
 - |\uparrow \downarrow \downarrow \uparrow \rangle
 + |\uparrow \downarrow \uparrow \downarrow \rangle
  ]   \right) \,.
\end{eqnarray}

In summary, the full spectrum of $\mathcal{H}_+$ consists of 186 states of vanishing energy, 16 states with energies $\pm U$ and $\pm \sqrt{2}U$ each, 2 modes for $\pm 2U$ and a unique state at energy $\pm 2\sqrt{2}U$. 

It is also interesting to notice that lowest excitation has an energy of $\omega=2(\sqrt{2}-1)U$ and changes the spin by $\Delta S^z=\pm1$. It is straight forward to show that the lowest states $\phi_0$ and $\phi_1$ are related as $[(S_a^+ - S_b^+) - (S_c^+ - S_d^+)] \phi_0 \sim \phi_{1} $. Note that the $b$ and $d$ sites have negative $\gamma_5$ eigenvalues. This, combined with the second relative minus sign between $(a,b)$ and $(c,d)$ indicates the relation to the operator $M^\beta_\mathbf{ i }$ defined in \eq{Eq:MFchannel} which has already been identified as the most dangerous mean-field channel. Here, we learn that this operator also exhibits the lowest excitation of the strongly interacting limit.

We have shown that the block $\mathcal{H}_+$ exhibit a gapped and unique ground state, which is therefore also symmetric. The overall ground state of the full lattice model is then a product state that is also gapped, unique and respects all symmetries.
Allowing finite, but small values of $t$ will change details of the ground-state wave-function. Especially it will no longer be a product state of purely local states. However, it will remain unique and symmetric until the hopping energy-scale set by $t$ is comparable to the many-body gap. This is quite opposite to local interactions with a degenerate ground state, e.g., the regular Hubbard model on the two-dimensional square lattice, where small perturbations can generate symmetry broken states, e.g., anti-ferromagnetic order due to a finite hopping amplitude in the Hubbard model.

To summarize, the limiting cases all exhibit unique ground states that share the same symmetries. As the strongly interacting state is a local product state, it is a representation of an atomic limit and very likely adiabatically connected to the trivial band insulators. The topological insulator is also stable against small interactions that were specifically chosen to allow a connection of the topological state to the strong interacting limit as the topologically protected defect states are symmetrically gapped \cite{queiroz16}. However, especially for intermediate coupling strength along the path from the topological insulator to the Mott phase at strong interacting, the energy scales mix and spontaneous symmetry breaking might occur. The \textit{dangerous} channel here is given by pseudo-magnetic order.

\section{Method}
\label{spt:method}

\begin{figure}
	\begin{center}
		\includegraphics[width=0.995\columnwidth]{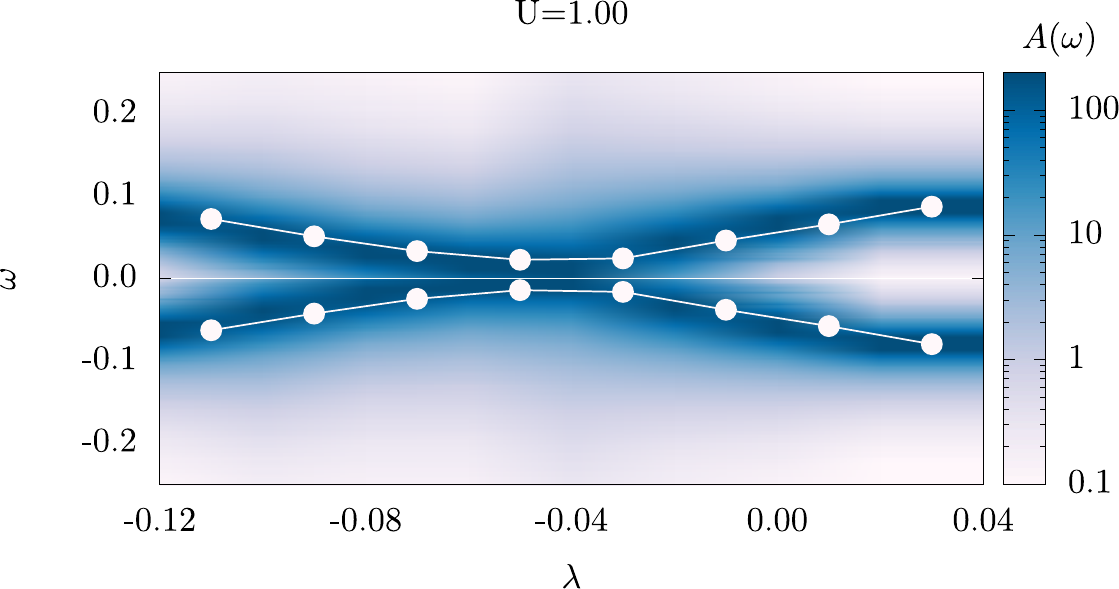}
	\end{center}
	\caption{\label{fig:sanity_check}Comparison of the single-particle spectrum $A(\omega)$ at the Dirac point $k=(\pi,\pi)$ extracted by analytic continuation using the stochastic maximum entropy method, with the band dispersion (solid black line) extracted from fitting the tail of time-displace Greens function. This proves that the assumption of a single low-energy excitation is justified.
	}
\end{figure}

\textit{Why should we go beyond mean field?} The considerations of \Sec{spt:MF} allow two scenarios for intermediate coupling strength, both are in agreement with the theorems on non-interacting systems. Either the order parameters vanishes and the bulk gap closes at the topological phase transition, or, the non-zero order parameters ensures a finite bulk gap at the expense of a broken symmetry. However the interaction constructed following the rules of \Ref{queiroz16} aims at a different path, namely an adiabatic connection of two topological distinct phases. Such a setup has to keep the band gap finite while preserving all protecting symmetries. Hence this connection cannot be made on a mean-field level and requires an intermediate state without a quasi particle description. In other words, one replaces poles of the Greens function that cross the Fermi surface (band gap closing) by zeros (no spectral weight) in order to change the topological invariant.

To solve the interacting system, we use the ALF package\cite{alf_v1}, a general implementation of the auxiliary field Quantum Monte Carlo\cite{blankenbecler81}. The zero-temperature version of this algorithm provides access to ground-state properties by using a trial wave function $|\psi_T\rangle$, we take the non-interacting ground state, and project it to the correlated one by applying the exponentiated Hamiltonian $\exp(-\Theta \mathcal{H})|\psi_T\rangle$ \cite{sugiyama86,sorella89}. Here $\Theta$ controls the projection length, the result converges exponentially fast and we choose $\Theta=20$ for the remainder of this work.
The implemented auxiliary-field QMC algorithm uses the Trotter decomposition of the partition sum $Z=\Tr{e^{-\beta(H_0+H_{\mathrm{int}})}}=\Tr{[e^{-\Delta \tau H_0} e^{-\Delta \tau  H_{\mathrm{int}} } ]^{ N_{\mathrm{Trotter}} } } + \mathcal{O}(\Delta\tau^2)$ with $\Delta \tau = \beta/N_{\mathrm{Trotter}}$ as well as a discrete Hubbard-Stratonovich transformation for the interaction of perfect squares $e^{\Delta\tau\hat{A}^2}=\sum_{l=\pm 1,\pm 2}\gamma(l)e^{\sqrt{\Delta\tau}\eta(l)\hat{A}} + \mathcal{O}(\Delta\tau^4)$. The Monte Carlo weight of each configuration $\{l_{\tau,i,\zeta,\alpha,\beta}\}$, where $\tau$ labels the imaginary time-slice and $i,\zeta,\alpha,\beta$ are the indices of \eq{Eq:HintSumSq}, is determined by integrating out the fermions and is given by the determinate of a matrix $W(\{l_{\tau,i,\zeta,\alpha,\beta}\})$.
   
The simulation of this model in the above formulation does not suffer a sign problem due an anti-unitary symmetry $\mathcal{T}$  defined as,    
\begin{equation}
\mathcal{T}^{-1}   \alpha
 \begin{bmatrix}
\Psi^{\dagger}_{\mathbf{i},A} \\
\Psi^{\dagger}_{\mathbf{i},B} \\
\Psi^{\dagger}_{\mathbf{i},C} \\
\Psi^{\dagger}_{\mathbf{i},D}  
\end{bmatrix}	
\mathcal{T}
=   \overline{\alpha} \begin{bmatrix}
 \Psi^{\dagger}_{\mathbf{i},D}\,\gamma_1 \gamma_5 \\
 \Psi^{\dagger}_{\mathbf{i},C}\,\gamma_1 \gamma_5 \\
 \Psi^{\dagger}_{\mathbf{i},B}\,\gamma_1 \gamma_5 \\
 \Psi^{\dagger}_{\mathbf{i},A}\,\gamma_1 \gamma_5  
\end{bmatrix}.
\end{equation} 
We note that $\gamma_1 \gamma_5$ commutes with the projection $P_\alpha$ and anti-commutes with $\gamma_5$ such that $\mathcal{T}^{-1} S_{\mathbf{i}}^{1,\alpha,\beta} \mathcal{T} = -S_{\mathbf{i}}^{2,\alpha,\beta}$. This symmetry is satisfied for each  field configuration and squares to $-\mathbf{1}$, such that the eigenvalues of $W(\{l_{\tau,i,\zeta,\alpha,\beta}\})$ come in complex conjugated pairs which guarantees the positivity of each configurations weight $\sim \det [W(\{l_{\tau,i,\zeta,\alpha,\beta}\})]$.  \cite{wu04}

Three observables are the main focus of this study. The first derivative of the free energy $\partial F / \partial U = -\beta/U\langle \mathcal{H}_{\mathrm{int}} \rangle$ signals a first-order phase transition for increasing interaction strength~$U$. To detect second-order phase transitions, we define a correlation ratio $r=1-\frac{S(\mathbf{q}=\delta\mathbf{q})}{S(\mathbf{q}=\mathbf{0})}$ where $S(\mathbf{q})$ is the correlation function in momentum space and $\delta\mathbf{q}$ the smallest but finite momentum available on the given lattice size. Observe that $\mathbf{q}=\mathbf{0}$ assumes a homogeneous instability which is justified according to the mean-field analysis. The correlation function is given as 
\begin{equation}
S(\mathbf{q})=\frac{1}{L^2} \sum_{\mathbf{i},\mathbf{j}} \exp [i\mathbf{q}(\mathbf{i}-\mathbf{j})] \sum_{\beta=x,y} \left\langle M^\beta_{\mathbf{i}} M^\beta_{\mathbf{j}} \right\rangle\,.
\end{equation}
In case of long-range homogeneous order $S(\mathbf{0})$ diverges linearly with system size $L^2$ whereas the correlation function remains finite for any other value of $\mathbf{q}$, hence $r=1$ for the thermodynamic limit. In systems without long-range order, $S(\mathbf{q})$ is a smooth function such that $r$ converges to zero for large lattices. As $r$ is an RG-invariant quantity it exhibits a crossing point for different system sizes at the phase transition.

The third observable of interest is the single-particle gap that allows us to track the semi-metallic Dirac cones that separate the insulators of different topology for $U=0$. The single-particle greens functions are given by $G_{\mathbf{k}}^p(\tau)=\sum_{o}\langle \mathcal{T} \Psi^{\phantom{\dagger}}_{\mathbf{k},o}(\tau)  \Psi^{\dagger}_{-\mathbf{k},o}(0) \rangle$ for particle excitations and $G_{\mathbf{k}}^h(\tau)=-\sum_{o}\langle \mathcal{T} \Psi^{\phantom{\dagger}}_{\mathbf{k},o}(0) \Psi^{\dagger}_{-\mathbf{k},o}(\tau) \rangle$ for hole excitations. If we assume a single quasi-particle mode at low energies gapped from higher energy excitations, then both greens function behave as $G_{\mathbf{k}}(\tau)\sim a_{\mathbf{k}} \exp(-\tau \epsilon_{\mathbf{k}})$ for large values of $\tau$ where $\epsilon_{\mathbf{k}}$ is the excitation energy and $a_{\mathbf{k}}$ its spectral weight.
As a sanity check of this assumption, we compare the extracted energies with the full spectrum (see \fig{fig:sanity_check}) determined by MaxEnt \cite{beach04a} which proves that the assumption is justified.

\section{Results} 
\label{spt:results}

We have shown above that the limit of strong interaction generates a gapped and symmetric ground state. Those two properties also apply to both non-interacting ground states, representing $-2<\lambda<0$ and $0<\lambda$, such that the adiabatic connection seems to be plausible, at least in principle. To test this hypothesis, we will track the semi-metallic phase and analyze the most dangerous correlation function identified in the mean-field considerations.

\begin{figure*}
	\begin{center}
		\includegraphics[width=0.99\textwidth]{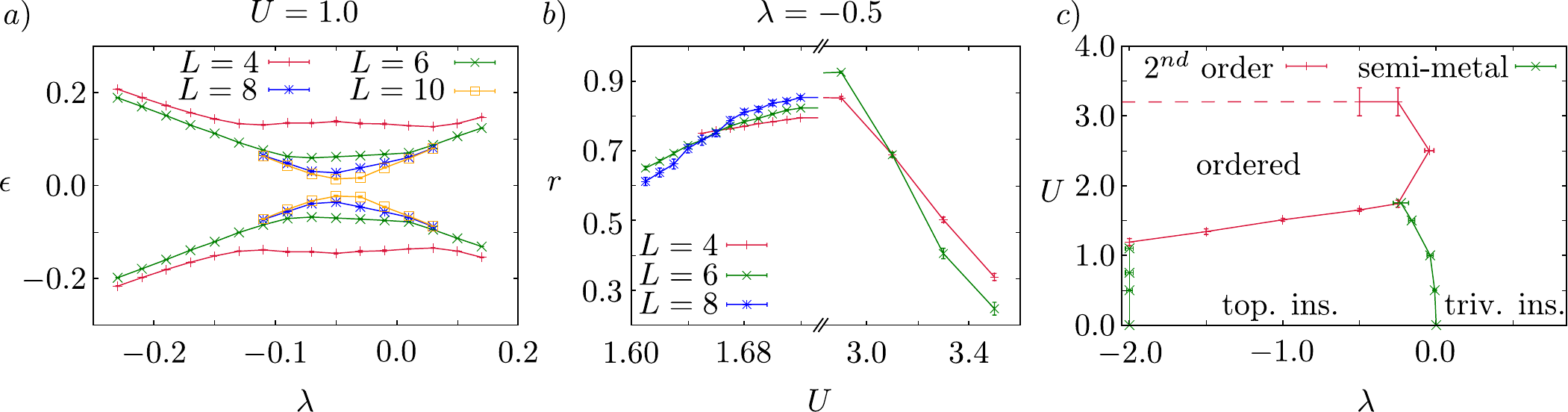}
	\end{center}
	\caption{\label{fig:PhaseDiagram}On the left hand side, we present the extracted energies $\epsilon_{\mathbf{k}=(\pi,\pi)}$ for the lowest particle/hole excitation. The system size scaling suggest a gap closing for $\lambda_c=-0.04\pm0.02$. In the central panel, the correlation ratio $r$ is presented and the size scaling is consistent with a symmetry broken phase between $U_c=1.65\pm0.02$ and $U_c=3.2\pm0.2$. The right hand side summarizes various scans in the phase diagram.
	}
\end{figure*}

\subsection{Tracking the semi-metal} Note that the two Hamiltonians with $\pm(\lambda+2)$ can be mapped onto each other. First, the mapping has to shift the momenta as $\mathbf{k}\rightarrow \mathbf{k}+(\pi,\pi)$ such that $m_{(\lambda+2)}(\mathbf{k})=-m_{-(\lambda+2)}(\mathbf{k}+(\pi,\pi))$. To absorb the sign changes in the first three terms of \eq{Eq:H_R_sector} the Dirac spinor has to transform as $\Psi_{\mathbf{k},o}\rightarrow \gamma_4\Psi_{\mathbf{k},o}$. Therefore, the position of the semi-metal with two Dirac cones at $(\pi,0)$ and $(0,\pi)$ has to remain at  fixed $\lambda=-2.0$ whereas the Dirac cone at $(\pi,\pi)$ generically occurs at renormalized values $\lambda\sim0$. It is interesting to notice that the $\gamma_4$ anti-commutes with $R$ such that the winding $w\rightarrow -w$ is inverted and the two Hamiltonian represent opposite topologies.

One might also be concerned that the interaction might lead to a meandering of the Dirac cone within the Brillouin zone. If we had kept the PHS with $\Delta=0$, then the cones are symmetry constraint to the time-reversal invariant momenta. On one hand, we fine-tuned the symmetry breaking such that the Dirac cones remain gapless in the free fermion system, and on the other hand, the numerical results show that the Dirac cone remain where they are.

Let us introduce $\lambda_c(U)$ as the critical value at which the semi-metal marks the topological phase transition between the TI with winding $w=+4$ and the trivial insulator. For the free fermion system, we have $\lambda_c(0)=0$. To locate the phase transition, we set a fixed interaction strength, e.g., $U=1.0$, and scan the single particle spectrum for various values of $\lambda$. The resulting excitation energies $\epsilon_{(\pi,\pi)}$ of the Dirac cone are presented in \fig{fig:PhaseDiagram}(a). The results depend on the lattice size and a visual extrapolation suggests a semi-metal  at $\lambda_c(1.0)=-0.04\pm0.02$. We repeat this analysis for various values of $U$ and also confirm the symmetry constrained position $\lambda_c(U)=-2.0$ for the Dirac semi-metal with cones at $(0,\pi)$ and $(\pi,0)$ which separates the $w=+4$ TI from the one with winding $w=-4$ at $-4.0\lesssim\lambda<-2.0$. The results are summarized in panel (c).

\subsection{Symmetry-broken phase} Here, we focus on the intermediate region of the phase diagram where the energy scales of the correlation and the kinetic energy compete with each other. During the mean-field analysis (see \Sec{spt:MF}), we have identified this regime between the TI and the Mott insulator at strong interactions as the one most prone to spontaneous symmetry breaking with long-range pseudo-magnetic order. 

Let us start with a fixed value of $\lambda=-0.5$ and analyze the correlation ratio $r$ with increasing interaction strength $U$. The resulting data is depicted in \fig{fig:PhaseDiagram}(b) for various lattice sizes. We clearly see that the ratio systematically decreases with increasing $L$ if the correlation strength is smaller than $U_c=1.65\pm0.02$ or larger than $U_c=3.2\pm0.2$ such that there is no long-range order here. In the intermediate region, the correlation ratio increases with system size that indicates spontaneous symmetry breaking due to a finite pseudo-magnetic order in $xy$-plane of $M_\mathbf{i}^\beta$. The critical values stated above are extracted from the crossing point where the ratio coincides for all lattices. The second phase transition from the ordered phase to the Mott insulator requires larger interaction strength such that the QMC simulation become more challenging, hence the smaller lattice sizes and larger error estimate.
Again, we repeat this calculation for various values of $\lambda$ and summarize the phase boundary in panel~(c).

\subsection{Phase diagram} In panel (c), we plot the full phase diagram and confirm the expected stability of the Dirac semi-metals as well as the insulators at weak coupling strength. The simulations also detect the symmetric state with strong correlation. In the middle of the phase diagram, where kinetic and potential energies are of similar order, we find long-range order in exactly the \textit{dangerous} channel that we have identified in \Sec{spt:MF}. This phase breaks the protecting $R$ symmetry and therefore allows a hybridization of counter-propagating edge modes as shown in \Fig{fig:MFoBC}. As a result, we cannot find an adiabatic path between the two non-interaction topological insulators. Instead, any path in this phase diagram either contains a semi-metallic state or a symmetry-breaking phase.

\begin{figure*}
	\begin{center}
		\includegraphics[width=0.99\textwidth]{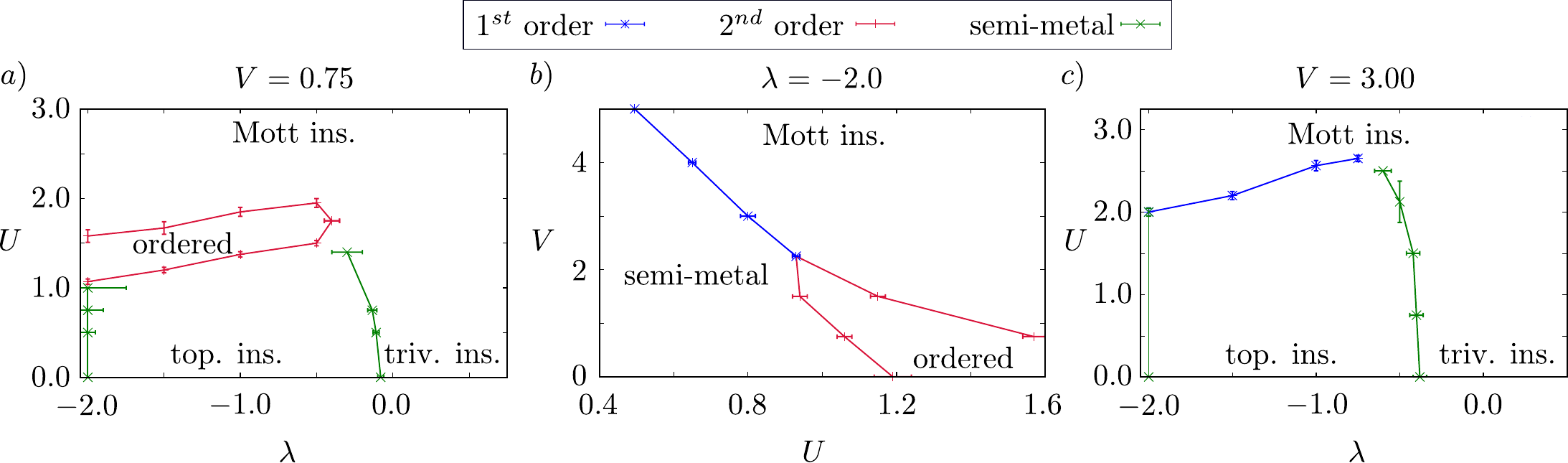}
	\end{center}
	\caption{\label{fig:frustPhaseDiagram}The left hand side show the phase diagram for weak frustration $V=0.75$ with a smaller region of long-range order compared to $V=0$ that seems to be most stable for $\lambda=-2$. In the middle, we present the phase diagram for fixed $\lambda=-2.0$ for various frustration strength $V$. For $V>2.5$, the symmetry broken phase is replaced by a first order phase transition. On the right hand side, the phase diagram is presented for high levels of frustration.
	}
\end{figure*}

\subsection{Can frustration remedy the problem?}
The main idea is to add the $z$-component of the pseudo-spins defined in \eq{eq:pseudospin} and use this to frustrate the in-plane order without changing the wave functions of the limiting cases. To form a proper $SU(2)$ algebra, we have to drop the $\gamma_5$ matrix acting on the Dirac components as $(\gamma_5)^2=\mathbf{1}$ such that
\begin{subequations}
	\label{eq:pseudospinZ}
	\begin{eqnarray}
	S_{\mathbf{i}}^{1,\alpha,z}&=& 
	(\Psi^{\dagger}_{\mathbf{i},A} \Psi^{\dagger}_{\mathbf{i},B}) 
	\mu_z  P_{\alpha}
%	\begin{pmatrix}
	(\Psi_{\mathbf{i},A} , \Psi_{\mathbf{i},B})^T
%	\end{pmatrix} 
	\label{Eq:PseudoSpinZ1} \\
	S_{\mathbf{i}}^{2,\alpha,z}&=& 
	(\Psi^{\dagger}_{\mathbf{i},C} \Psi^{\dagger}_{\mathbf{i},D}) 
	\mu_z  P_{\alpha}
%	\begin{pmatrix}
	( \Psi_{\mathbf{i},C}, \Psi_{\mathbf{i},D})^T
%	\end{pmatrix}
	\,. \label{Eq:PseudoSpinZ2}
	\end{eqnarray}
\end{subequations}
Observe that this $z$-component generates rotations within the $xy$-plane of the pseudo-spins that leave the Hamiltonian invariant. Additionally, the transformation $(A\leftrightarrow B)$ combined with $(C\leftrightarrow D)$ also is a symmetry operation under which $S_{\mathbf{i}}^{1/2,\alpha,z}\rightarrow -S_{\mathbf{i}}^{1/2,\alpha,z}$ is inverted. Hence, any unique ground state has to be an eigenstate of $\sum_{\mathbf{i},\alpha=\pm} \left( S_{\mathbf{i}}^{1,\alpha,z} +  \, S_{\mathbf{i}}^{2,\alpha,z} \right)$ with a vanishing expectation value. In the large $U$ limit, the sites and $\alpha$-sub-blocks decouple such that we introduce an additional interaction term $\mathcal{H}_{\mathrm{frust}}=V\sum_{\mathbf{i},\alpha=\pm} \left( S_{\mathbf{i}}^{1,\alpha,z} +  \, S_{\mathbf{i}}^{2,\alpha,z} \right)^2$ which minimizes $S^z$ locally without changing the ground state.

As depicted in \fig{fig:frustPhaseDiagram}, weak frustration does reduce the size of the symmetry broken phase. In panel (a), we present the phase diagram for $V=0.75$. The symmetry broken phase now extends only to $\lambda\sim0.5$ whereas in the unfrustrated model it reaches $\lambda\sim0$. Additionally, we find that the long-range ordered region is shifted towards weaker coupling strength $U$ while also the range in $U$ has been reduced. This trend is also clearly visible in panel (b) where we kept $\lambda=-2.0$ fixed. The Dirac cones at $(\pi,0)$ and $(0,\pi)$ persist for weak coupling strength $U$ and $V$. Increasing $U$ generates the long-ranged ordered state before the appearance of the symmetric Mott insulator at large $U$. With higher level of frustration the symmetry broken phase is replaced by a direct first order phase transition between the Dirac semi-metal and the Mott insulator.
In \fig{fig:frustPhaseDiagram}(c), we show that the first order phase transition extends also to $\lambda>-2.0$ and connects to other semi-metal of the $(\pi,\pi)$ Dirac cone. 

\section{Discussion} 
\label{spt:discussion}

In this study, we found an interaction which at sufficient strength trivializes the topological phase $w = 4$. At the same time, surprisingly, it does not allow for an adiabatic path between this topological phase and the trivial phase $w = 0$.
The semi-metal separating the non-interacting insulators persists for small coupling strength in $U$. It is terminated either by a first order transition to a symmetric Mott insulator related to the large $U$ limit or a second-order phase transition to a long-range ordered phase. This in turn is separated by another second-order transition at larger $U$ to the symmetric Mott insulator. Similarly,  the topological insulator either undergoes a direct first order transition to the Mott phase or a second order transition into a symmetry-broken phase followed by another transition into the symmetric Mott phase.

We emphasize that our results do not contradict the statement that the non-interacting classification is reduced from $\Z$ to $\Z_4$ with interaction. Instead, our  results show that the existence of a specific interactions which symmetrically gaps out the surface states of a topological insulator is not sufficient to establish that the corresponding bulk phase can be adiabatically connected to a trivial phase. This contradicts a popular line of reasoning which focuses on gapping out the surface states as a criterion for establishing the classification reduction. The underlying reason for the failure of such arguments is that the stability of surface states can be essentially reduced to a zero-dimensional problem by studying the zero-energy states within defects constructed in a specific way \cite{queiroz16}.
However, this decreased dimensionality does have strong implications on the possibility of spontaneous symmetry breaking in the ground state that may only occur in one (two) or higher dimensions for discrete (continuous) symmetries according to Mermin-Wagner theorem \cite{mermin66}. And it is exactly this mechanism which blocks the path we were looking for.

From the two-dimensional bulk perspective, the phase diagram exhibits various critical points. Even though the focus of this study was  to establish the phases themselves, it is interesting to discuss those critical theories briefly. There are Wilson-Fisher transitions between the topological insulator and the ordered phase as well as between the ordered phase and the Mott insulator. Additionally, we expect the critical point where the semi-metal is gapped by symmetry breaking to be described by a Gross-Neveu theory\cite{herbut09a,toldin14}.

\begin{figure}
	\begin{center}
		\includegraphics[width=0.995\columnwidth]{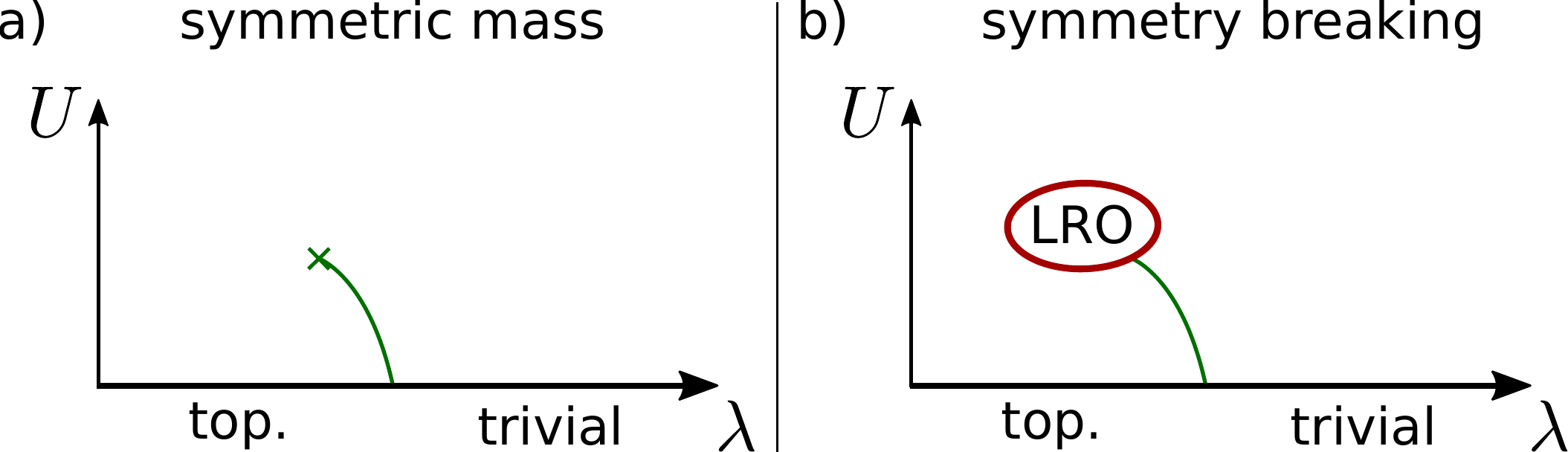}
	\end{center}
	\caption{\label{fig:dis}There exist at least two alternative bulk scenarios with (a) symmetric mass generation (green cross) or (b) a symmetry breaking region (red circle) with long-range order (LRO) terminating the semi-metal between topologically distinct non-interacting states. Observe that the region with LRO connects to only one semi-metal and not two both as in our numerical phase diagram in \Fig{fig:PhaseDiagram}(c). Both scenarios allow an adiabatic path and are not realized in the range of parameters investigated in this study.
	}
\end{figure}

The results of this work raise a few questions, mainly focused on the missing pieces required to find the adiabatic path. In \fig{fig:dis}, we sketch two alternative scenarios for the bulk phase diagram, (a) the symmetric mass generation for the Dirac cone as well as (b) a separated region of symmetry breaking that terminates the semi-metal line. Several studies have reported the formation of a correlated single-particle gap of $SU(4)$-symmetric Dirac system without the generation of long-range order \cite{ayyar15,He16}. Most surprisingly, it is claimed to be a second order transition. In \Ref{you18}, the authors propose a theory that involves fractionalization in order to explain this exotic phase transition. It would be very promising to include the same kind of bulk criticality in our setup in order to find the adiabatic connection and then investigate the details of how this affects the topological aspects. In contrast, the scenario (b) shows that this symmetric mass generation is not required and that there also exist more conventional options in which the symmetry broken region we find is split into two separate ones.

However, there is no obvious approach to engineer this phase diagram. One possibility is to consider the symmetry of the interaction. Upon using highly symmetric interactions in the flavor space, the previous studies \cite{he17} found a direct second-order transition to the symmetric Mott phase suggesting the possibility of an adiabatic path between the topological and trivial phases. In contrast, the interaction we use has very low symmetry in the flavor space, and we always found a symmetry-broken phase or a first order transition which completely blocks such adiabatic path. It would be interesting to see if interactions which are more symmetric than ours but less symmetric than those considered previously could yield a phase diagram similar to Fig.~\ref{fig:dis}a).

Another possibility yet again involves the bulk-boundary correspondence, similar to previous studies, in order to fine-tune the energy scales of the model. 
The energy scale of the helical edge state is set by its Fermi velocity $v_\mathrm{edge}$. Interestingly, it is possible to reduce $v_\mathrm{edge}$ without a significant change of the bulk gap\cite{budich14}. This is achieved by localizing the Berry curvature in momentum space. Hence, we can control and therefore separate the bulk energy scale $\Delta_\mathrm{bulk}$ from the edge $v_\mathrm{edge}$ without introducing open boundary condition by using bulk parameters. This constitutes a promising avenue towards the aforementioned hierarchy of energy scales $v_\mathrm{edge}<U<\Delta_\mathrm{bulk}$, and this hierarchy is also often used in the arguments on the reduced topological classifications. There, the interaction is supposed to be strong enough to gap out the edge spectrum and therefore trivializes the topology, but still weak enough to avoid the broken symmetry in the bulk. Here, it is usually believed that the phase transition of the edge coincides with the topological phase transition of the bulk and thus enables the adiabatic connection.

However, this also demonstrates an important subtlety of bulk boundary correspondence in such systems. In particular, what would happen if we consider an interaction which is very weak compared to the bulk energy scale $U/ \Delta_{\rm bulk} \ll 1$ so that we expect the phase to be adiabatically connected a non-interacting bulk topological phase while at the same time being rather strong compared to the characteristic energy scale at the edge $U/v_\mathrm{edge}>U_\mathrm{edge}^c/v_\mathrm{edge}$ (where $U_\mathrm{edge}^c$ the critical interaction strength). Thus, the bulk and edge phase transition do not have to coincide, at least based on the involved energy scales.
There are at least two distinct scenarios in this case: (i) the edge may gap out by spontanuously breaking the protecting symmetry, (ii) it may gap-out symmetrically and become topologically trivial \footnote{In higher dimensions, there is also a possibility that the edge symmetrically gaps out developping topological order}.
Whereas the latter scenario violates bulk boundary correspondence, the former does not. In this case, we expect the symmetry at the edge to be restored as the strength of the interaction is increased beyond a characteristic scale associated with the \emph{bulk} rather than the \emph{edge}. Investigating which of these scenarios is realized represents a very interesting extension of the current work.

\begin{acknowledgments}
The authors thank Erez Berg, Akira Furusaki, Cenke Xu, and Yizhuang You for stimulating discussions. JH  was supported by the German Research Foundation (DFG), under DFG-SFB~1170 ``ToCoTronics" (Project C01).  FFA acknowledges financial support from the DFG through the W\"urzburg-Dresden Cluster of Excellence on Complexity and Topology in Quantum Matter - ct.qmat (EXC 2147, project-id 39085490).
EK was supported by a Simons Investigator Fellowship, by NSF-DMR 1411343, and by the German National Academy of Sciences Leopoldina through grant LPDS 2018-02 Leopoldina fellowship. RQ was funded by the  Deutsche Forschungsgemeinschaft (DFG, German Research
Foundation) Projektnummer 277101999 TRR 183 (project
B03), the Israel Science Foundation, and the European Research
Council (Project LEGOTOP).
 
The authors gratefully acknowledge the Gauss Centre for Supercomputing e.V. (www.gauss-centre.eu) for funding this project by providing computing time on the GCS Supercomputer SuperMUC at Leibniz Supercomputing Centre (LRZ, www.lrz.de). The authors also gratefully acknowledge the computing time granted by the John von Neumann Institute for Computing (NIC) and provided on the supercomputer JURECA at J\"ulich Supercomputing Centre (JSC).
\end{acknowledgments}

\bibliography{newbib.bib}

\begin{thebibliography}{56}%
\makeatletter
\providecommand \@ifxundefined [1]{%
 \@ifx{#1\undefined}
}%
\providecommand \@ifnum [1]{%
 \ifnum #1\expandafter \@firstoftwo
 \else \expandafter \@secondoftwo
 \fi
}%
\providecommand \@ifx [1]{%
 \ifx #1\expandafter \@firstoftwo
 \else \expandafter \@secondoftwo
 \fi
}%
\providecommand \natexlab [1]{#1}%
\providecommand \enquote  [1]{``#1''}%
\providecommand \bibnamefont  [1]{#1}%
\providecommand \bibfnamefont [1]{#1}%
\providecommand \citenamefont [1]{#1}%
\providecommand \href@noop [0]{\@secondoftwo}%
\providecommand \href [0]{\begingroup \@sanitize@url \@href}%
\providecommand \@href[1]{\@@startlink{#1}\@@href}%
\providecommand \@@href[1]{\endgroup#1\@@endlink}%
\providecommand \@sanitize@url [0]{\catcode `\\12\catcode `\$12\catcode
  `\&12\catcode `\#12\catcode `\^12\catcode `\_12\catcode `\%12\relax}%
\providecommand \@@startlink[1]{}%
\providecommand \@@endlink[0]{}%
\providecommand \url  [0]{\begingroup\@sanitize@url \@url }%
\providecommand \@url [1]{\endgroup\@href {#1}{\urlprefix }}%
\providecommand \urlprefix  [0]{URL }%
\providecommand \Eprint [0]{\href }%
\providecommand \doibase [0]{http://dx.doi.org/}%
\providecommand \selectlanguage [0]{\@gobble}%
\providecommand \bibinfo  [0]{\@secondoftwo}%
\providecommand \bibfield  [0]{\@secondoftwo}%
\providecommand \translation [1]{[#1]}%
\providecommand \BibitemOpen [0]{}%
\providecommand \bibitemStop [0]{}%
\providecommand \bibitemNoStop [0]{.\EOS\space}%
\providecommand \EOS [0]{\spacefactor3000\relax}%
\providecommand \BibitemShut  [1]{\csname bibitem#1\endcsname}%
\let\auto@bib@innerbib\@empty
%</preamble>
\bibitem [{\citenamefont {Hasan}\ and\ \citenamefont
  {Kane}(2010)}]{hasan10_rev}%
  \BibitemOpen
  \bibfield  {author} {\bibinfo {author} {\bibfnamefont {M.~Z.}\ \bibnamefont
  {Hasan}}\ and\ \bibinfo {author} {\bibfnamefont {C.~L.}\ \bibnamefont
  {Kane}},\ }\href {\doibase 10.1103/RevModPhys.82.3045} {\bibfield  {journal}
  {\bibinfo  {journal} {Rev. Mod. Phys.}\ }\textbf {\bibinfo {volume} {82}},\
  \bibinfo {pages} {3045} (\bibinfo {year} {2010})}\BibitemShut {NoStop}%
\bibitem [{\citenamefont {Qi}\ and\ \citenamefont {Zhang}(2011)}]{qi11}%
  \BibitemOpen
  \bibfield  {author} {\bibinfo {author} {\bibfnamefont {X.-L.}\ \bibnamefont
  {Qi}}\ and\ \bibinfo {author} {\bibfnamefont {S.-C.}\ \bibnamefont {Zhang}},\
  }\href {\doibase 10.1103/RevModPhys.83.1057} {\bibfield  {journal} {\bibinfo
  {journal} {Rev. Mod. Phys.}\ }\textbf {\bibinfo {volume} {83}},\ \bibinfo
  {pages} {1057} (\bibinfo {year} {2011})}\BibitemShut {NoStop}%
\bibitem [{\citenamefont {Chiu}\ \emph {et~al.}(2016)\citenamefont {Chiu},
  \citenamefont {Teo}, \citenamefont {Schnyder},\ and\ \citenamefont
  {Ryu}}]{chiu_review15}%
  \BibitemOpen
  \bibfield  {author} {\bibinfo {author} {\bibfnamefont {C.-K.}\ \bibnamefont
  {Chiu}}, \bibinfo {author} {\bibfnamefont {J.~C.~Y.}\ \bibnamefont {Teo}},
  \bibinfo {author} {\bibfnamefont {A.~P.}\ \bibnamefont {Schnyder}}, \ and\
  \bibinfo {author} {\bibfnamefont {S.}~\bibnamefont {Ryu}},\ }\href {\doibase
  10.1103/RevModPhys.88.035005} {\bibfield  {journal} {\bibinfo  {journal}
  {Rev. Mod. Phys.}\ }\textbf {\bibinfo {volume} {88}},\ \bibinfo {pages}
  {035005} (\bibinfo {year} {2016})}\BibitemShut {NoStop}%
\bibitem [{\citenamefont {Ludwig}(2015)}]{ludwig15}%
  \BibitemOpen
  \bibfield  {author} {\bibinfo {author} {\bibfnamefont {A.~W.~W.}\
  \bibnamefont {Ludwig}},\ }\href {\doibase 10.1088/0031-8949/2015/t168/014001}
  {\bibfield  {journal} {\bibinfo  {journal} {Physica Scripta}\ }\textbf
  {\bibinfo {volume} {T168}},\ \bibinfo {pages} {014001} (\bibinfo {year}
  {2015})}\BibitemShut {NoStop}%
\bibitem [{\citenamefont {Schnyder}\ \emph {et~al.}(2008)\citenamefont
  {Schnyder}, \citenamefont {Ryu}, \citenamefont {Furusaki},\ and\
  \citenamefont {Ludwig}}]{schnyder08}%
  \BibitemOpen
  \bibfield  {author} {\bibinfo {author} {\bibfnamefont {A.~P.}\ \bibnamefont
  {Schnyder}}, \bibinfo {author} {\bibfnamefont {S.}~\bibnamefont {Ryu}},
  \bibinfo {author} {\bibfnamefont {A.}~\bibnamefont {Furusaki}}, \ and\
  \bibinfo {author} {\bibfnamefont {A.~W.~W.}\ \bibnamefont {Ludwig}},\ }\href
  {\doibase 10.1103/PhysRevB.78.195125} {\bibfield  {journal} {\bibinfo
  {journal} {Phys. Rev. B}\ }\textbf {\bibinfo {volume} {78}},\ \bibinfo
  {pages} {195125} (\bibinfo {year} {2008})}\BibitemShut {NoStop}%
\bibitem [{\citenamefont {Ryu}\ \emph {et~al.}(2010)\citenamefont {Ryu},
  \citenamefont {Schnyder}, \citenamefont {Furusaki},\ and\ \citenamefont
  {Ludwig}}]{ryu10}%
  \BibitemOpen
  \bibfield  {author} {\bibinfo {author} {\bibfnamefont {S.}~\bibnamefont
  {Ryu}}, \bibinfo {author} {\bibfnamefont {A.~P.}\ \bibnamefont {Schnyder}},
  \bibinfo {author} {\bibfnamefont {A.}~\bibnamefont {Furusaki}}, \ and\
  \bibinfo {author} {\bibfnamefont {A.~W.~W.}\ \bibnamefont {Ludwig}},\ }\href
  {http://stacks.iop.org/1367-2630/12/i=6/a=065010} {\bibfield  {journal}
  {\bibinfo  {journal} {New Journal of Physics}\ }\textbf {\bibinfo {volume}
  {12}},\ \bibinfo {pages} {065010} (\bibinfo {year} {2010})}\BibitemShut
  {NoStop}%
\bibitem [{\citenamefont {{Tsui}}\ \emph {et~al.}(1982)\citenamefont {{Tsui}},
  \citenamefont {{Stormer}},\ and\ \citenamefont {{Gossard}}}]{Tsui82}%
  \BibitemOpen
  \bibfield  {author} {\bibinfo {author} {\bibfnamefont {D.~C.}\ \bibnamefont
  {{Tsui}}}, \bibinfo {author} {\bibfnamefont {H.~L.}\ \bibnamefont
  {{Stormer}}}, \ and\ \bibinfo {author} {\bibfnamefont {A.~C.}\ \bibnamefont
  {{Gossard}}},\ }\href {\doibase 10.1103/PhysRevLett.48.1559} {\bibfield
  {journal} {\bibinfo  {journal} {Physical Review Letters}\ }\textbf {\bibinfo
  {volume} {48}},\ \bibinfo {pages} {1559} (\bibinfo {year}
  {1982})}\BibitemShut {NoStop}%
\bibitem [{\citenamefont {Stormer}\ \emph {et~al.}(1999)\citenamefont
  {Stormer}, \citenamefont {Tsui},\ and\ \citenamefont {Gossard}}]{stromer99}%
  \BibitemOpen
  \bibfield  {author} {\bibinfo {author} {\bibfnamefont {H.~L.}\ \bibnamefont
  {Stormer}}, \bibinfo {author} {\bibfnamefont {D.~C.}\ \bibnamefont {Tsui}}, \
  and\ \bibinfo {author} {\bibfnamefont {A.~C.}\ \bibnamefont {Gossard}},\
  }\href {\doibase 10.1103/RevModPhys.71.S298} {\bibfield  {journal} {\bibinfo
  {journal} {Rev. Mod. Phys.}\ }\textbf {\bibinfo {volume} {71}},\ \bibinfo
  {pages} {S298} (\bibinfo {year} {1999})}\BibitemShut {NoStop}%
\bibitem [{\citenamefont {Levin}\ and\ \citenamefont {Stern}(2009)}]{levin09}%
  \BibitemOpen
  \bibfield  {author} {\bibinfo {author} {\bibfnamefont {M.}~\bibnamefont
  {Levin}}\ and\ \bibinfo {author} {\bibfnamefont {A.}~\bibnamefont {Stern}},\
  }\href {\doibase 10.1103/PhysRevLett.103.196803} {\bibfield  {journal}
  {\bibinfo  {journal} {Phys. Rev. Lett.}\ }\textbf {\bibinfo {volume} {103}},\
  \bibinfo {pages} {196803} (\bibinfo {year} {2009})}\BibitemShut {NoStop}%
\bibitem [{\citenamefont {{Balents}}(2010)}]{balents-10}%
  \BibitemOpen
  \bibfield  {author} {\bibinfo {author} {\bibfnamefont {L.}~\bibnamefont
  {{Balents}}},\ }\href {\doibase 10.1038/nature08917} {\bibfield  {journal}
  {\bibinfo  {journal} {\nat}\ }\textbf {\bibinfo {volume} {464}},\ \bibinfo
  {pages} {199} (\bibinfo {year} {2010})}\BibitemShut {NoStop}%
\bibitem [{\citenamefont {Lee}(2008)}]{Lee08}%
  \BibitemOpen
  \bibfield  {author} {\bibinfo {author} {\bibfnamefont {P.~A.}\ \bibnamefont
  {Lee}},\ }\href {\doibase 10.1126/science.1163196} {\bibfield  {journal}
  {\bibinfo  {journal} {Science}\ }\textbf {\bibinfo {volume} {321}},\ \bibinfo
  {pages} {1306} (\bibinfo {year} {2008})},\ \Eprint
  {http://arxiv.org/abs/https://science.sciencemag.org/content/321/5894/1306.full.pdf}
  {https://science.sciencemag.org/content/321/5894/1306.full.pdf} \BibitemShut
  {NoStop}%
\bibitem [{\citenamefont {Yamashita}\ \emph {et~al.}(2010)\citenamefont
  {Yamashita}, \citenamefont {Nakata}, \citenamefont {Senshu}, \citenamefont
  {Nagata}, \citenamefont {Yamamoto}, \citenamefont {Kato}, \citenamefont
  {Shibauchi},\ and\ \citenamefont {Matsuda}}]{Yamashita10}%
  \BibitemOpen
  \bibfield  {author} {\bibinfo {author} {\bibfnamefont {M.}~\bibnamefont
  {Yamashita}}, \bibinfo {author} {\bibfnamefont {N.}~\bibnamefont {Nakata}},
  \bibinfo {author} {\bibfnamefont {Y.}~\bibnamefont {Senshu}}, \bibinfo
  {author} {\bibfnamefont {M.}~\bibnamefont {Nagata}}, \bibinfo {author}
  {\bibfnamefont {H.~M.}\ \bibnamefont {Yamamoto}}, \bibinfo {author}
  {\bibfnamefont {R.}~\bibnamefont {Kato}}, \bibinfo {author} {\bibfnamefont
  {T.}~\bibnamefont {Shibauchi}}, \ and\ \bibinfo {author} {\bibfnamefont
  {Y.}~\bibnamefont {Matsuda}},\ }\href {\doibase 10.1126/science.1188200}
  {\bibfield  {journal} {\bibinfo  {journal} {Science}\ }\textbf {\bibinfo
  {volume} {328}},\ \bibinfo {pages} {1246} (\bibinfo {year} {2010})},\ \Eprint
  {http://arxiv.org/abs/https://science.sciencemag.org/content/328/5983/1246.full.pdf}
  {https://science.sciencemag.org/content/328/5983/1246.full.pdf} \BibitemShut
  {NoStop}%
\bibitem [{\citenamefont {{Sachdev}}(2008)}]{sachdev08}%
  \BibitemOpen
  \bibfield  {author} {\bibinfo {author} {\bibfnamefont {S.}~\bibnamefont
  {{Sachdev}}},\ }\href {\doibase 10.1038/nphys894} {\bibfield  {journal}
  {\bibinfo  {journal} {Nature Physics}\ }\textbf {\bibinfo {volume} {4}},\
  \bibinfo {pages} {173} (\bibinfo {year} {2008})},\ \Eprint
  {http://arxiv.org/abs/0711.3015} {arXiv:0711.3015 [cond-mat.str-el]}
  \BibitemShut {NoStop}%
\bibitem [{\citenamefont {Fidkowski}\ and\ \citenamefont
  {Kitaev}(2010)}]{fidkowski10a}%
  \BibitemOpen
  \bibfield  {author} {\bibinfo {author} {\bibfnamefont {L.}~\bibnamefont
  {Fidkowski}}\ and\ \bibinfo {author} {\bibfnamefont {A.}~\bibnamefont
  {Kitaev}},\ }\href {\doibase 10.1103/PhysRevB.81.134509} {\bibfield
  {journal} {\bibinfo  {journal} {Phys. Rev. B}\ }\textbf {\bibinfo {volume}
  {81}},\ \bibinfo {pages} {134509} (\bibinfo {year} {2010})}\BibitemShut
  {NoStop}%
\bibitem [{\citenamefont {Tang}\ and\ \citenamefont {Wen}(2012)}]{tang12}%
  \BibitemOpen
  \bibfield  {author} {\bibinfo {author} {\bibfnamefont {E.}~\bibnamefont
  {Tang}}\ and\ \bibinfo {author} {\bibfnamefont {X.-G.}\ \bibnamefont {Wen}},\
  }\href {\doibase 10.1103/PhysRevLett.109.096403} {\bibfield  {journal}
  {\bibinfo  {journal} {Phys. Rev. Lett.}\ }\textbf {\bibinfo {volume} {109}},\
  \bibinfo {pages} {096403} (\bibinfo {year} {2012})}\BibitemShut {NoStop}%
\bibitem [{\citenamefont {Yao}\ and\ \citenamefont {Ryu}(2013)}]{yao13}%
  \BibitemOpen
  \bibfield  {author} {\bibinfo {author} {\bibfnamefont {H.}~\bibnamefont
  {Yao}}\ and\ \bibinfo {author} {\bibfnamefont {S.}~\bibnamefont {Ryu}},\
  }\href {\doibase 10.1103/PhysRevB.88.064507} {\bibfield  {journal} {\bibinfo
  {journal} {Phys. Rev. B}\ }\textbf {\bibinfo {volume} {88}},\ \bibinfo
  {pages} {064507} (\bibinfo {year} {2013})}\BibitemShut {NoStop}%
\bibitem [{\citenamefont {Morimoto}\ \emph {et~al.}(2015)\citenamefont
  {Morimoto}, \citenamefont {Furusaki},\ and\ \citenamefont
  {Mudry}}]{morimoto15}%
  \BibitemOpen
  \bibfield  {author} {\bibinfo {author} {\bibfnamefont {T.}~\bibnamefont
  {Morimoto}}, \bibinfo {author} {\bibfnamefont {A.}~\bibnamefont {Furusaki}},
  \ and\ \bibinfo {author} {\bibfnamefont {C.}~\bibnamefont {Mudry}},\ }\href
  {\doibase 10.1103/PhysRevB.92.125104} {\bibfield  {journal} {\bibinfo
  {journal} {Phys. Rev. B}\ }\textbf {\bibinfo {volume} {92}},\ \bibinfo
  {pages} {125104} (\bibinfo {year} {2015})}\BibitemShut {NoStop}%
\bibitem [{\citenamefont {Qi}(2013)}]{qi13}%
  \BibitemOpen
  \bibfield  {author} {\bibinfo {author} {\bibfnamefont {X.-L.}\ \bibnamefont
  {Qi}},\ }\href {\doibase 10.1088/1367-2630/15/6/065002} {\bibfield  {journal}
  {\bibinfo  {journal} {New Journal of Physics}\ }\textbf {\bibinfo {volume}
  {15}},\ \bibinfo {pages} {065002} (\bibinfo {year} {2013})}\BibitemShut
  {NoStop}%
\bibitem [{\citenamefont {Queiroz}\ \emph {et~al.}(2016)\citenamefont
  {Queiroz}, \citenamefont {Khalaf},\ and\ \citenamefont {Stern}}]{queiroz16}%
  \BibitemOpen
  \bibfield  {author} {\bibinfo {author} {\bibfnamefont {R.}~\bibnamefont
  {Queiroz}}, \bibinfo {author} {\bibfnamefont {E.}~\bibnamefont {Khalaf}}, \
  and\ \bibinfo {author} {\bibfnamefont {A.}~\bibnamefont {Stern}},\ }\href
  {\doibase 10.1103/PhysRevLett.117.206405} {\bibfield  {journal} {\bibinfo
  {journal} {Phys. Rev. Lett.}\ }\textbf {\bibinfo {volume} {117}},\ \bibinfo
  {pages} {206405} (\bibinfo {year} {2016})}\BibitemShut {NoStop}%
\bibitem [{\citenamefont {Turner}\ \emph {et~al.}(2011)\citenamefont {Turner},
  \citenamefont {Pollmann},\ and\ \citenamefont {Berg}}]{turner11}%
  \BibitemOpen
  \bibfield  {author} {\bibinfo {author} {\bibfnamefont {A.~M.}\ \bibnamefont
  {Turner}}, \bibinfo {author} {\bibfnamefont {F.}~\bibnamefont {Pollmann}}, \
  and\ \bibinfo {author} {\bibfnamefont {E.}~\bibnamefont {Berg}},\ }\href
  {\doibase 10.1103/PhysRevB.83.075102} {\bibfield  {journal} {\bibinfo
  {journal} {Phys. Rev. B}\ }\textbf {\bibinfo {volume} {83}},\ \bibinfo
  {pages} {075102} (\bibinfo {year} {2011})}\BibitemShut {NoStop}%
\bibitem [{\citenamefont {Fidkowski}\ and\ \citenamefont
  {Kitaev}(2011)}]{fidkowski11}%
  \BibitemOpen
  \bibfield  {author} {\bibinfo {author} {\bibfnamefont {L.}~\bibnamefont
  {Fidkowski}}\ and\ \bibinfo {author} {\bibfnamefont {A.}~\bibnamefont
  {Kitaev}},\ }\href {\doibase 10.1103/PhysRevB.83.075103} {\bibfield
  {journal} {\bibinfo  {journal} {Phys. Rev. B}\ }\textbf {\bibinfo {volume}
  {83}},\ \bibinfo {pages} {075103} (\bibinfo {year} {2011})}\BibitemShut
  {NoStop}%
\bibitem [{\citenamefont {Gu}\ and\ \citenamefont {Levin}(2014)}]{gu14}%
  \BibitemOpen
  \bibfield  {author} {\bibinfo {author} {\bibfnamefont {Z.-C.}\ \bibnamefont
  {Gu}}\ and\ \bibinfo {author} {\bibfnamefont {M.}~\bibnamefont {Levin}},\
  }\href {\doibase 10.1103/PhysRevB.89.201113} {\bibfield  {journal} {\bibinfo
  {journal} {Phys. Rev. B}\ }\textbf {\bibinfo {volume} {89}},\ \bibinfo
  {pages} {201113} (\bibinfo {year} {2014})}\BibitemShut {NoStop}%
\bibitem [{\citenamefont {Wang}\ and\ \citenamefont {Senthil}(2014)}]{wang14e}%
  \BibitemOpen
  \bibfield  {author} {\bibinfo {author} {\bibfnamefont {C.}~\bibnamefont
  {Wang}}\ and\ \bibinfo {author} {\bibfnamefont {T.}~\bibnamefont {Senthil}},\
  }\href {\doibase 10.1103/PhysRevB.89.195124} {\bibfield  {journal} {\bibinfo
  {journal} {Phys. Rev. B}\ }\textbf {\bibinfo {volume} {89}},\ \bibinfo
  {pages} {195124} (\bibinfo {year} {2014})}\BibitemShut {NoStop}%
\bibitem [{\citenamefont {Senthil}(2015)}]{senthil15}%
  \BibitemOpen
  \bibfield  {author} {\bibinfo {author} {\bibfnamefont {T.}~\bibnamefont
  {Senthil}},\ }\href {\doibase 10.1146/annurev-conmatphys-031214-014740}
  {\bibfield  {journal} {\bibinfo  {journal} {Annual Review of Condensed Matter
  Physics}\ }\textbf {\bibinfo {volume} {6}},\ \bibinfo {pages} {299} (\bibinfo
  {year} {2015})},\ \Eprint
  {http://arxiv.org/abs/https://doi.org/10.1146/annurev-conmatphys-031214-014740}
  {https://doi.org/10.1146/annurev-conmatphys-031214-014740} \BibitemShut
  {NoStop}%
\bibitem [{\citenamefont {Vishwanath}\ and\ \citenamefont
  {Senthil}(2013)}]{vishwanath13}%
  \BibitemOpen
  \bibfield  {author} {\bibinfo {author} {\bibfnamefont {A.}~\bibnamefont
  {Vishwanath}}\ and\ \bibinfo {author} {\bibfnamefont {T.}~\bibnamefont
  {Senthil}},\ }\href {\doibase 10.1103/PhysRevX.3.011016} {\bibfield
  {journal} {\bibinfo  {journal} {Phys. Rev. X}\ }\textbf {\bibinfo {volume}
  {3}},\ \bibinfo {pages} {011016} (\bibinfo {year} {2013})}\BibitemShut
  {NoStop}%
\bibitem [{\citenamefont {He}\ \emph {et~al.}(2016)\citenamefont {He},
  \citenamefont {Wu}, \citenamefont {You}, \citenamefont {Xu}, \citenamefont
  {Meng},\ and\ \citenamefont {Lu}}]{He16}%
  \BibitemOpen
  \bibfield  {author} {\bibinfo {author} {\bibfnamefont {Y.-Y.}\ \bibnamefont
  {He}}, \bibinfo {author} {\bibfnamefont {H.-Q.}\ \bibnamefont {Wu}}, \bibinfo
  {author} {\bibfnamefont {Y.-Z.}\ \bibnamefont {You}}, \bibinfo {author}
  {\bibfnamefont {C.}~\bibnamefont {Xu}}, \bibinfo {author} {\bibfnamefont
  {Z.~Y.}\ \bibnamefont {Meng}}, \ and\ \bibinfo {author} {\bibfnamefont
  {Z.-Y.}\ \bibnamefont {Lu}},\ }\href {\doibase 10.1103/PhysRevB.94.241111}
  {\bibfield  {journal} {\bibinfo  {journal} {Phys. Rev. B}\ }\textbf {\bibinfo
  {volume} {94}},\ \bibinfo {pages} {241111} (\bibinfo {year}
  {2016})}\BibitemShut {NoStop}%
\bibitem [{\citenamefont {Ayyar}\ and\ \citenamefont
  {Chandrasekharan}(2016)}]{ayyar15}%
  \BibitemOpen
  \bibfield  {author} {\bibinfo {author} {\bibfnamefont {V.}~\bibnamefont
  {Ayyar}}\ and\ \bibinfo {author} {\bibfnamefont {S.}~\bibnamefont
  {Chandrasekharan}},\ }\href {\doibase 10.1103/PhysRevD.93.081701} {\bibfield
  {journal} {\bibinfo  {journal} {Phys. Rev. D}\ }\textbf {\bibinfo {volume}
  {93}},\ \bibinfo {pages} {081701} (\bibinfo {year} {2016})}\BibitemShut
  {NoStop}%
\bibitem [{\citenamefont {Blankenbecler}\ \emph {et~al.}(1981)\citenamefont
  {Blankenbecler}, \citenamefont {Scalapino},\ and\ \citenamefont
  {Sugar}}]{blankenbecler81}%
  \BibitemOpen
  \bibfield  {author} {\bibinfo {author} {\bibfnamefont {R.}~\bibnamefont
  {Blankenbecler}}, \bibinfo {author} {\bibfnamefont {D.~J.}\ \bibnamefont
  {Scalapino}}, \ and\ \bibinfo {author} {\bibfnamefont {R.~L.}\ \bibnamefont
  {Sugar}},\ }\href {\doibase 10.1103/PhysRevD.24.2278} {\bibfield  {journal}
  {\bibinfo  {journal} {Phys. Rev. D}\ }\textbf {\bibinfo {volume} {24}},\
  \bibinfo {pages} {2278} (\bibinfo {year} {1981})}\BibitemShut {NoStop}%
\bibitem [{\citenamefont {White}\ \emph {et~al.}(1989)\citenamefont {White},
  \citenamefont {Scalapino}, \citenamefont {Sugar}, \citenamefont {Loh},
  \citenamefont {Gubernatis},\ and\ \citenamefont {Scalettar}}]{white89}%
  \BibitemOpen
  \bibfield  {author} {\bibinfo {author} {\bibfnamefont {S.}~\bibnamefont
  {White}}, \bibinfo {author} {\bibfnamefont {D.}~\bibnamefont {Scalapino}},
  \bibinfo {author} {\bibfnamefont {R.}~\bibnamefont {Sugar}}, \bibinfo
  {author} {\bibfnamefont {E.}~\bibnamefont {Loh}}, \bibinfo {author}
  {\bibfnamefont {J.}~\bibnamefont {Gubernatis}}, \ and\ \bibinfo {author}
  {\bibfnamefont {R.}~\bibnamefont {Scalettar}},\ }\href {\doibase
  10.1103/PhysRevB.40.506} {\bibfield  {journal} {\bibinfo  {journal} {Phys.
  Rev. B}\ }\textbf {\bibinfo {volume} {40}},\ \bibinfo {pages} {506} (\bibinfo
  {year} {1989})}\BibitemShut {NoStop}%
\bibitem [{\citenamefont {Assaad}\ and\ \citenamefont
  {Evertz}(2008)}]{assaad08_rev}%
  \BibitemOpen
  \bibfield  {author} {\bibinfo {author} {\bibfnamefont {F.}~\bibnamefont
  {Assaad}}\ and\ \bibinfo {author} {\bibfnamefont {H.}~\bibnamefont
  {Evertz}},\ }in\ \href {\doibase 10.1007/978-3-540-74686-7_10} {\emph
  {\bibinfo {booktitle} {Computational Many-Particle Physics}}},\ \bibinfo
  {series} {Lecture Notes in Physics}, Vol.\ \bibinfo {volume} {739},\ \bibinfo
  {editor} {edited by\ \bibinfo {editor} {\bibfnamefont {H.}~\bibnamefont
  {Fehske}}, \bibinfo {editor} {\bibfnamefont {R.}~\bibnamefont {Schneider}}, \
  and\ \bibinfo {editor} {\bibfnamefont {A.}~\bibnamefont {Wei{\ss}e}}}\
  (\bibinfo  {publisher} {Springer},\ \bibinfo {address} {Berlin Heidelberg},\
  \bibinfo {year} {2008})\ pp.\ \bibinfo {pages} {277--356}\BibitemShut
  {NoStop}%
\bibitem [{\citenamefont {Bercx}\ \emph {et~al.}(2017)\citenamefont {Bercx},
  \citenamefont {Goth}, \citenamefont {Hofmann},\ and\ \citenamefont
  {Assaad}}]{alf_v1}%
  \BibitemOpen
  \bibfield  {author} {\bibinfo {author} {\bibfnamefont {M.}~\bibnamefont
  {Bercx}}, \bibinfo {author} {\bibfnamefont {F.}~\bibnamefont {Goth}},
  \bibinfo {author} {\bibfnamefont {J.~S.}\ \bibnamefont {Hofmann}}, \ and\
  \bibinfo {author} {\bibfnamefont {F.~F.}\ \bibnamefont {Assaad}},\ }\href
  {\doibase 10.21468/SciPostPhys.3.2.013} {\bibfield  {journal} {\bibinfo
  {journal} {SciPost Phys.}\ }\textbf {\bibinfo {volume} {3}},\ \bibinfo
  {pages} {013} (\bibinfo {year} {2017})}\BibitemShut {NoStop}%
\bibitem [{\citenamefont {{Kane}}\ and\ \citenamefont
  {{Mele}}(2005)}]{kane2005}%
  \BibitemOpen
  \bibfield  {author} {\bibinfo {author} {\bibfnamefont {C.~L.}\ \bibnamefont
  {{Kane}}}\ and\ \bibinfo {author} {\bibfnamefont {E.~J.}\ \bibnamefont
  {{Mele}}},\ }\href {\doibase 10.1103/PhysRevLett.95.146802} {\bibfield
  {journal} {\bibinfo  {journal} {\prl}\ }\textbf {\bibinfo {volume} {95}},\
  \bibinfo {eid} {146802} (\bibinfo {year} {2005})},\ \Eprint
  {http://arxiv.org/abs/cond-mat/0506581} {cond-mat/0506581} \BibitemShut
  {NoStop}%
\bibitem [{\citenamefont {{Bernevig}}\ \emph {et~al.}(2006)\citenamefont
  {{Bernevig}}, \citenamefont {{Hughes}},\ and\ \citenamefont
  {{Zhang}}}]{bernevig2006}%
  \BibitemOpen
  \bibfield  {author} {\bibinfo {author} {\bibfnamefont {B.~A.}\ \bibnamefont
  {{Bernevig}}}, \bibinfo {author} {\bibfnamefont {T.~L.}\ \bibnamefont
  {{Hughes}}}, \ and\ \bibinfo {author} {\bibfnamefont {S.-C.}\ \bibnamefont
  {{Zhang}}},\ }\href {\doibase 10.1126/science.1133734} {\bibfield  {journal}
  {\bibinfo  {journal} {Science}\ }\textbf {\bibinfo {volume} {314}},\ \bibinfo
  {pages} {1757} (\bibinfo {year} {2006})},\ \Eprint
  {http://arxiv.org/abs/cond-mat/0611399} {cond-mat/0611399} \BibitemShut
  {NoStop}%
\bibitem [{\citenamefont {Qi}\ \emph {et~al.}(2008)\citenamefont {Qi},
  \citenamefont {Hughes},\ and\ \citenamefont {Zhang}}]{qi08a}%
  \BibitemOpen
  \bibfield  {author} {\bibinfo {author} {\bibfnamefont {X.-L.}\ \bibnamefont
  {Qi}}, \bibinfo {author} {\bibfnamefont {T.~L.}\ \bibnamefont {Hughes}}, \
  and\ \bibinfo {author} {\bibfnamefont {S.-C.}\ \bibnamefont {Zhang}},\ }\href
  {\doibase 10.1103/PhysRevB.78.195424} {\bibfield  {journal} {\bibinfo
  {journal} {Phys. Rev. B}\ }\textbf {\bibinfo {volume} {78}},\ \bibinfo
  {pages} {195424} (\bibinfo {year} {2008})}\BibitemShut {NoStop}%
\bibitem [{\citenamefont {Liu}\ \emph {et~al.}(2008)\citenamefont {Liu},
  \citenamefont {Qi}, \citenamefont {Dai}, \citenamefont {Fang},\ and\
  \citenamefont {Zhang}}]{liu08}%
  \BibitemOpen
  \bibfield  {author} {\bibinfo {author} {\bibfnamefont {C.-X.}\ \bibnamefont
  {Liu}}, \bibinfo {author} {\bibfnamefont {X.-L.}\ \bibnamefont {Qi}},
  \bibinfo {author} {\bibfnamefont {X.}~\bibnamefont {Dai}}, \bibinfo {author}
  {\bibfnamefont {Z.}~\bibnamefont {Fang}}, \ and\ \bibinfo {author}
  {\bibfnamefont {S.-C.}\ \bibnamefont {Zhang}},\ }\href {\doibase
  10.1103/PhysRevLett.101.146802} {\bibfield  {journal} {\bibinfo  {journal}
  {Phys. Rev. Lett.}\ }\textbf {\bibinfo {volume} {101}},\ \bibinfo {pages}
  {146802} (\bibinfo {year} {2008})}\BibitemShut {NoStop}%
\bibitem [{\citenamefont {Haldane}(1988)}]{haldane88a}%
  \BibitemOpen
  \bibfield  {author} {\bibinfo {author} {\bibfnamefont {F.~D.~M.}\
  \bibnamefont {Haldane}},\ }\href {\doibase 10.1103/PhysRevLett.61.2015}
  {\bibfield  {journal} {\bibinfo  {journal} {Phys. Rev. Lett.}\ }\textbf
  {\bibinfo {volume} {61}},\ \bibinfo {pages} {2015} (\bibinfo {year}
  {1988})}\BibitemShut {NoStop}%
\bibitem [{Note1()}]{Note1}%
  \BibitemOpen
  \bibinfo {note} {Note that due to the broken particle-hole symmetry, the
  topological states do not necessarily have zero energy. However, a
  renormalized chemical potential can only shift the energy but it cannot lift
  their degeneracy.}\BibitemShut {Stop}%
\bibitem [{Note2()}]{Note2}%
  \BibitemOpen
  \bibinfo {note} {The attentive reader might have realized that this
  correlated hopping actually changes $R$ by $4$. However, in the canonical
  ensemble, the total number of particles is conserved and at half-filling with
  an even number of fermions, the total $R$ charge itself has to be even and
  thus this reduces the $U(1)$ symmetry to $\protect \mathbb
  {Z}_2$.}\BibitemShut {Stop}%
\bibitem [{\citenamefont {Sorella}\ and\ \citenamefont
  {Tosatti}(1992)}]{sorella92}%
  \BibitemOpen
  \bibfield  {author} {\bibinfo {author} {\bibfnamefont {S.}~\bibnamefont
  {Sorella}}\ and\ \bibinfo {author} {\bibfnamefont {E.}~\bibnamefont
  {Tosatti}},\ }\href@noop {} {\bibfield  {journal} {\bibinfo  {journal}
  {Europhys. Lett.}\ }\textbf {\bibinfo {volume} {19}},\ \bibinfo {pages} {699}
  (\bibinfo {year} {1992})}\BibitemShut {NoStop}%
\bibitem [{\citenamefont {Paiva}\ \emph {et~al.}(2005)\citenamefont {Paiva},
  \citenamefont {Scalettar}, \citenamefont {Zheng}, \citenamefont {Singh},\
  and\ \citenamefont {Oitmaa}}]{paiva05}%
  \BibitemOpen
  \bibfield  {author} {\bibinfo {author} {\bibfnamefont {T.}~\bibnamefont
  {Paiva}}, \bibinfo {author} {\bibfnamefont {R.~T.}\ \bibnamefont
  {Scalettar}}, \bibinfo {author} {\bibfnamefont {W.}~\bibnamefont {Zheng}},
  \bibinfo {author} {\bibfnamefont {R.~R.~P.}\ \bibnamefont {Singh}}, \ and\
  \bibinfo {author} {\bibfnamefont {J.}~\bibnamefont {Oitmaa}},\ }\href@noop {}
  {\bibfield  {journal} {\bibinfo  {journal} {Phys. Rev. B}\ }\textbf {\bibinfo
  {volume} {72}},\ \bibinfo {pages} {085123} (\bibinfo {year}
  {2005})}\BibitemShut {NoStop}%
\bibitem [{\citenamefont {Meng}\ \emph {et~al.}(2010)\citenamefont {Meng},
  \citenamefont {Lang}, \citenamefont {Wessel}, \citenamefont {Assaad},\ and\
  \citenamefont {Muramatsu}}]{meng10}%
  \BibitemOpen
  \bibfield  {author} {\bibinfo {author} {\bibfnamefont {Z.~Y.}\ \bibnamefont
  {Meng}}, \bibinfo {author} {\bibfnamefont {T.~C.}\ \bibnamefont {Lang}},
  \bibinfo {author} {\bibfnamefont {S.}~\bibnamefont {Wessel}}, \bibinfo
  {author} {\bibfnamefont {F.~F.}\ \bibnamefont {Assaad}}, \ and\ \bibinfo
  {author} {\bibfnamefont {A.}~\bibnamefont {Muramatsu}},\ }\href
  {http://dx.doi.org/10.1038/nature08942} {\bibfield  {journal} {\bibinfo
  {journal} {Nature}\ }\textbf {\bibinfo {volume} {464}},\ \bibinfo {pages}
  {847} (\bibinfo {year} {2010})}\BibitemShut {NoStop}%
\bibitem [{\citenamefont {Sorella}\ \emph {et~al.}(2012)\citenamefont
  {Sorella}, \citenamefont {Otsuka},\ and\ \citenamefont {Yunoki}}]{sorella12}%
  \BibitemOpen
  \bibfield  {author} {\bibinfo {author} {\bibfnamefont {S.}~\bibnamefont
  {Sorella}}, \bibinfo {author} {\bibfnamefont {Y.}~\bibnamefont {Otsuka}}, \
  and\ \bibinfo {author} {\bibfnamefont {S.}~\bibnamefont {Yunoki}},\ }\href
  {\doibase http://dx.doi.org/10.1038/srep00992} {\bibfield  {journal}
  {\bibinfo  {journal} {Sci. Rep.}\ }\textbf {\bibinfo {volume} {2}},\ \bibinfo
  {pages} {992} (\bibinfo {year} {2012})}\BibitemShut {NoStop}%
\bibitem [{\citenamefont {Assaad}\ and\ \citenamefont
  {Herbut}(2013)}]{assaad13}%
  \BibitemOpen
  \bibfield  {author} {\bibinfo {author} {\bibfnamefont {F.~F.}\ \bibnamefont
  {Assaad}}\ and\ \bibinfo {author} {\bibfnamefont {I.~F.}\ \bibnamefont
  {Herbut}},\ }\href {\doibase 10.1103/PhysRevX.3.031010} {\bibfield  {journal}
  {\bibinfo  {journal} {Phys. Rev. X}\ }\textbf {\bibinfo {volume} {3}},\
  \bibinfo {pages} {031010} (\bibinfo {year} {2013})}\BibitemShut {NoStop}%
\bibitem [{\citenamefont {Parisen~Toldin}\ \emph {et~al.}(2015)\citenamefont
  {Parisen~Toldin}, \citenamefont {Hohenadler}, \citenamefont {Assaad},\ and\
  \citenamefont {Herbut}}]{toldin14}%
  \BibitemOpen
  \bibfield  {author} {\bibinfo {author} {\bibfnamefont {F.}~\bibnamefont
  {Parisen~Toldin}}, \bibinfo {author} {\bibfnamefont {M.}~\bibnamefont
  {Hohenadler}}, \bibinfo {author} {\bibfnamefont {F.~F.}\ \bibnamefont
  {Assaad}}, \ and\ \bibinfo {author} {\bibfnamefont {I.~F.}\ \bibnamefont
  {Herbut}},\ }\href {\doibase 10.1103/PhysRevB.91.165108} {\bibfield
  {journal} {\bibinfo  {journal} {Phys. Rev. B}\ }\textbf {\bibinfo {volume}
  {91}},\ \bibinfo {pages} {165108} (\bibinfo {year} {2015})}\BibitemShut
  {NoStop}%
\bibitem [{Note3()}]{Note3}%
  \BibitemOpen
  \bibinfo {note} {Note that $\gamma _5$ and $P_\alpha $ commute with each
  other such that they may be diagonalized simultaneously. However, $\gamma _5$
  anti-commutes with $R$ such that there is no basis which diagonalizes all
  three of them.}\BibitemShut {Stop}%
\bibitem [{Note4()}]{Note4}%
  \BibitemOpen
  \bibinfo {note} {Actually, the local number of particles is conserved.
  However, in the following it is more useful to distinguish the parity only as
  both empty and double-occupied sites constitute a spin-singlet which
  effectively removes that site from the ring.}\BibitemShut {Stop}%
\bibitem [{\citenamefont {Sugiyama}\ and\ \citenamefont
  {Koonin}(1986)}]{sugiyama86}%
  \BibitemOpen
  \bibfield  {author} {\bibinfo {author} {\bibfnamefont {G.}~\bibnamefont
  {Sugiyama}}\ and\ \bibinfo {author} {\bibfnamefont {S.}~\bibnamefont
  {Koonin}},\ }\href {\doibase http://dx.doi.org/10.1016/0003-4916(86)90107-7}
  {\bibfield  {journal} {\bibinfo  {journal} {Annals of Physics}\ }\textbf
  {\bibinfo {volume} {168}},\ \bibinfo {pages} {1 } (\bibinfo {year}
  {1986})}\BibitemShut {NoStop}%
\bibitem [{\citenamefont {Sorella}\ \emph {et~al.}(1989)\citenamefont
  {Sorella}, \citenamefont {Baroni}, \citenamefont {Car},\ and\ \citenamefont
  {Parrinello}}]{sorella89}%
  \BibitemOpen
  \bibfield  {author} {\bibinfo {author} {\bibfnamefont {S.}~\bibnamefont
  {Sorella}}, \bibinfo {author} {\bibfnamefont {S.}~\bibnamefont {Baroni}},
  \bibinfo {author} {\bibfnamefont {R.}~\bibnamefont {Car}}, \ and\ \bibinfo
  {author} {\bibfnamefont {M.}~\bibnamefont {Parrinello}},\ }\href
  {http://stacks.iop.org/0295-5075/8/i=7/a=014} {\bibfield  {journal} {\bibinfo
   {journal} {EPL (Europhysics Letters)}\ }\textbf {\bibinfo {volume} {8}},\
  \bibinfo {pages} {663} (\bibinfo {year} {1989})}\BibitemShut {NoStop}%
\bibitem [{\citenamefont {Wu}\ and\ \citenamefont {Zhang}(2005)}]{wu04}%
  \BibitemOpen
  \bibfield  {author} {\bibinfo {author} {\bibfnamefont {C.}~\bibnamefont
  {Wu}}\ and\ \bibinfo {author} {\bibfnamefont {S.-C.}\ \bibnamefont {Zhang}},\
  }\href {\doibase 10.1103/PhysRevB.71.155115} {\bibfield  {journal} {\bibinfo
  {journal} {Phys. Rev. B}\ }\textbf {\bibinfo {volume} {71}},\ \bibinfo
  {pages} {155115} (\bibinfo {year} {2005})}\BibitemShut {NoStop}%
\bibitem [{\citenamefont {{Beach}}(2004)}]{beach04a}%
  \BibitemOpen
  \bibfield  {author} {\bibinfo {author} {\bibfnamefont {K.~S.~D.}\
  \bibnamefont {{Beach}}},\ }\href@noop {} {\bibfield  {journal} {\bibinfo
  {journal} {eprint arXiv:cond-mat/0403055}\ } (\bibinfo {year} {2004})},\
  \Eprint {http://arxiv.org/abs/cond-mat/0403055} {cond-mat/0403055}
  \BibitemShut {NoStop}%
\bibitem [{\citenamefont {Mermin}\ and\ \citenamefont
  {Wagner}(1966)}]{mermin66}%
  \BibitemOpen
  \bibfield  {author} {\bibinfo {author} {\bibfnamefont {N.~D.}\ \bibnamefont
  {Mermin}}\ and\ \bibinfo {author} {\bibfnamefont {H.}~\bibnamefont
  {Wagner}},\ }\href {\doibase 10.1103/PhysRevLett.17.1133} {\bibfield
  {journal} {\bibinfo  {journal} {Phys. Rev. Lett.}\ }\textbf {\bibinfo
  {volume} {17}},\ \bibinfo {pages} {1133} (\bibinfo {year}
  {1966})}\BibitemShut {NoStop}%
\bibitem [{\citenamefont {Herbut}\ \emph {et~al.}(2009)\citenamefont {Herbut},
  \citenamefont {Juri\ifmmode \check{c}\else \v{c}\fi{}i\ifmmode~\acute{c}\else
  \'{c}\fi{}},\ and\ \citenamefont {Vafek}}]{herbut09a}%
  \BibitemOpen
  \bibfield  {author} {\bibinfo {author} {\bibfnamefont {I.~F.}\ \bibnamefont
  {Herbut}}, \bibinfo {author} {\bibfnamefont {V.}~\bibnamefont {Juri\ifmmode
  \check{c}\else \v{c}\fi{}i\ifmmode~\acute{c}\else \'{c}\fi{}}}, \ and\
  \bibinfo {author} {\bibfnamefont {O.}~\bibnamefont {Vafek}},\ }\href
  {\doibase 10.1103/PhysRevB.80.075432} {\bibfield  {journal} {\bibinfo
  {journal} {Phys. Rev. B}\ }\textbf {\bibinfo {volume} {80}},\ \bibinfo
  {pages} {075432} (\bibinfo {year} {2009})}\BibitemShut {NoStop}%
\bibitem [{\citenamefont {You}\ \emph {et~al.}(2018)\citenamefont {You},
  \citenamefont {He}, \citenamefont {Xu},\ and\ \citenamefont
  {Vishwanath}}]{you18}%
  \BibitemOpen
  \bibfield  {author} {\bibinfo {author} {\bibfnamefont {Y.-Z.}\ \bibnamefont
  {You}}, \bibinfo {author} {\bibfnamefont {Y.-C.}\ \bibnamefont {He}},
  \bibinfo {author} {\bibfnamefont {C.}~\bibnamefont {Xu}}, \ and\ \bibinfo
  {author} {\bibfnamefont {A.}~\bibnamefont {Vishwanath}},\ }\href {\doibase
  10.1103/PhysRevX.8.011026} {\bibfield  {journal} {\bibinfo  {journal} {Phys.
  Rev. X}\ }\textbf {\bibinfo {volume} {8}},\ \bibinfo {pages} {011026}
  (\bibinfo {year} {2018})}\BibitemShut {NoStop}%
\bibitem [{\citenamefont {He}\ \emph {et~al.}(2018)\citenamefont {He},
  \citenamefont {Xu}, \citenamefont {Sun}, \citenamefont {Assaad},
  \citenamefont {Meng},\ and\ \citenamefont {Lu}}]{he17}%
  \BibitemOpen
  \bibfield  {author} {\bibinfo {author} {\bibfnamefont {Y.-Y.}\ \bibnamefont
  {He}}, \bibinfo {author} {\bibfnamefont {X.~Y.}\ \bibnamefont {Xu}}, \bibinfo
  {author} {\bibfnamefont {K.}~\bibnamefont {Sun}}, \bibinfo {author}
  {\bibfnamefont {F.~F.}\ \bibnamefont {Assaad}}, \bibinfo {author}
  {\bibfnamefont {Z.~Y.}\ \bibnamefont {Meng}}, \ and\ \bibinfo {author}
  {\bibfnamefont {Z.-Y.}\ \bibnamefont {Lu}},\ }\href {\doibase
  10.1103/PhysRevB.97.081110} {\bibfield  {journal} {\bibinfo  {journal} {Phys.
  Rev. B}\ }\textbf {\bibinfo {volume} {97}},\ \bibinfo {pages} {081110}
  (\bibinfo {year} {2018})}\BibitemShut {NoStop}%
\bibitem [{\citenamefont {Budich}\ \emph {et~al.}(2014)\citenamefont {Budich},
  \citenamefont {Eisert},\ and\ \citenamefont {Bergholtz}}]{budich14}%
  \BibitemOpen
  \bibfield  {author} {\bibinfo {author} {\bibfnamefont {J.~C.}\ \bibnamefont
  {Budich}}, \bibinfo {author} {\bibfnamefont {J.}~\bibnamefont {Eisert}}, \
  and\ \bibinfo {author} {\bibfnamefont {E.~J.}\ \bibnamefont {Bergholtz}},\
  }\href {\doibase 10.1103/PhysRevB.89.195120} {\bibfield  {journal} {\bibinfo
  {journal} {Phys. Rev. B}\ }\textbf {\bibinfo {volume} {89}},\ \bibinfo
  {pages} {195120} (\bibinfo {year} {2014})}\BibitemShut {NoStop}%
\bibitem [{Note5()}]{Note5}%
  \BibitemOpen
  \bibinfo {note} {In higher dimensions, there is also a possibility that the
  edge symmetrically gaps out developping topological order}\BibitemShut
  {NoStop}%
\end{thebibliography}%

\end{document}